\documentclass[12pt]{article}

 \usepackage[left=1in,
right=1in,
top=1in,
bottom=1.2in,
footskip=.3in]{geometry}

\usepackage{amsmath}
\usepackage{placeins}
\usepackage{amsfonts}
\usepackage{amssymb}
\usepackage{hyperref}
\usepackage{xcolor} % coloring text
\usepackage{setspace}
\allowdisplaybreaks %  allow page breaks inside equations

\usepackage{graphicx,subcaption}
\graphicspath{{./Examples/}} 
\usepackage{amsfonts,yfonts}
\usepackage{algorithm,algpseudocode}

\usepackage{pifont}% http://ctan.org/pkg/pifont
\usepackage{pgf, tikz}
\usetikzlibrary{arrows, automata}
\usetikzlibrary{graphs,graphs.standard,calc}
\usepackage{tikz-qtree}
\usepackage{arydshln, booktabs}

\newtheorem{theorem}{Theorem}[section]

\newtheorem{fact}{Fact}[section]

\newtheorem{definition}{Definition}[section]
\newtheorem{example}{Example}[section]
\newtheorem{lemma}{Lemma}[section]
\newtheorem{summary}{Summary}[section]
\newtheorem{proposition}{Proposition}[section]
\newtheorem{remark}{Remark}[section]
\newenvironment{proof}[1][Proof]{\noindent\textbf{#1.} }{\ \rule{0.5em}{0.5em}}

\DeclareMathOperator{\bS}{\pmb{S}}

\DeclareMathOperator{\bY}{\pmb{Y}}
\DeclareMathOperator{\bZ}{\pmb{Z}}
\DeclareMathOperator{\bL}{\pmb{L}}

\DeclareMathOperator{\bN}{\pmb{N}}
\DeclareMathOperator{\bB}{\pmb{B}}
\DeclareMathOperator{\bC}{\pmb{C}}
\DeclareMathOperator{\bD}{\pmb{D}}
\DeclareMathOperator{\bmu}{\pmb{\mu}}
\DeclareMathOperator{\bnu}{\pmb{\nu}}
\DeclareMathOperator{\b1}{\pmb{1}}
\DeclareMathOperator{\bzero}{\pmb{0}}

\title{Modeling cyclicality and intransitivity in paired comparisons data}

\author{Rahul Singh$^1$ and Ori Davidov$^2$}

\date{$^1${Department of Mathematics, Indian Institute of Technology Delhi, Delhi 110016, India}\\
{$^2$Department of Statistics, University of Haifa, Mount Carmel, Haifa 3498838 Israel}\\
{\small
E-mail: \texttt{wrahulsingh@gmail.com} (R Singh), 
\texttt{davidov@stat.haifa.ac.il} (O Davidov)}}

\begin{document}

\maketitle
{\setstretch{1.2}
\begin{abstract} 
Paired comparison data arise in ranking problems, decision analysis, sports analytics, recommendation systems, and many other applications in which alternatives are evaluated by comparing two items at a time. Standard models typically impose a transitive preference profile induced by a vector of merits. In many empirical settings, however, preference relations exhibit cyclic and intransitive patterns that cannot be adequately represented by a global ranking. This paper develops a framework for modeling cyclicality and departures from transitivity. The proposed approach decomposes a preference profile into orthogonal transitive and cyclic components and provides a geometric characterization of the associated parameter space. The cyclic component is represented using an overcomplete dictionary of elementary cycles, so that identifying cyclic structure and the intransitivities it may induce becomes a sparse model selection problem. We propose a method for recovering sparse cyclic structure and establish large--sample guarantees for estimation and model recovery. The analysis clarifies the relationship between cyclicality, intransitivity, and several notions of transitivity used in paired comparison theory. By explicitly modeling cyclic structure, the proposed framework can improve estimation, ranking, interpretation, and prediction. The methodology is evaluated through simulations and illustrated with an empirical application.

\medskip


\end{abstract}
}
%%%%%%%%%%%%%%%%%%%%%%%%%%%%
%%%%%%%%%%%%%%%%%%%%%%%%%%%%
%%%%%%%%%%%%%%%%%%%%%%%%%%%%

\section{Introduction} \label{section:introduction}

Paired comparison data (PCD) arise in classical domains such as economics and psychology, as well as in operational settings such as decision analysis, sports analytics, recommendation systems, and ranking problems, in which preferences among a collection of items are elicited or observed through pairwise comparisons (David, 1988). Standard approaches typically impose transitivity, leading to models in which preferences are represented using latent scores or merits from which a global ranking is induced (Marden, 1995; Langville and Meyer, 2012). Such models play a central role in ranking, recommendation (Agarwal and Chen, 2016), and decision analysis. 
This is particularly evident in methods based on pairwise comparison matrices, such as the analytic hierarchy process (Saaty, 1977; Kou et al., 2016; Bozóki et al., 2016). In these settings, departures from consistency are commonly assessed through inconsistency indices. Since consistency is equivalent to transitivity in this context, sufficiently large departures are often reduced by revising the comparisons or by replacing the observed matrix with a nearby consistent approximation.

While convenient and widely used, merit--based models impose strong structural restrictions that may fail to capture important features of the data. Preference relations need not be transitive, and cyclic patterns, in which preferences form closed loops, are well documented. They arise in familiar examples such as rock--paper--scissors and Efron's intransitive dice (Gardner, 1970), and, more importantly, have been reported in empirical studies of choice and preference (Tversky, 1969; Birnbaum and Schmidt, 2008; Cavagnaro and Davis--Stober, 2014; Alós--Ferrer and Garagnani, 2021). Although the prevalence of such violations has been debated (Iverson and Falmagne, 1985; Loomes et al., 1992), intransitivity remains a recurring empirical phenomenon, underscoring the need for models that explicitly accommodate cyclic structure and the departures from transitivity it may induce.

We consider $K$ items labeled $1,\ldots,K$, and let $Y_{ijk}$ denote the outcome of the $k^{th}$ comparison between items $i$ and $j$. The random variable (RV) $Y_{ijk}$ may be binary, ordinal, or cardinal. In this paper, we focus on cardinal, or real--valued, PCD; however, the applicability of the proposed methodology to binary PCD is also discussed in Section \ref{discussion}. Specifically, we assume that the observations $Y_{ijk}$, for $1\leq i\neq j\leq K$ and $k=1,\ldots,n_{ij}$, satisfy
\begin{equation}
\label{model.Y_ijk}
Y_{ijk} = \nu_{ij} + \epsilon_{ijk},
\end{equation}
where $\nu_{ij}=\mathbb{E}(Y_{ijk})$ and the errors $\epsilon_{ijk}$ are independent and identically distributed random variables with zero mean and variance $\sigma^2$. The additive formulation also covers reciprocal pairwise comparison matrices, since logarithmic transformations are commonly used to represent multiplicative comparisons on an additive scale; see, e.g., Koczkodaj and Orlowski (1997). Let $\mathcal{G}=(\mathcal{V},\mathcal{E})$ denote the graph whose vertices are the items $1,\ldots,K$ and whose edges are the pairs $(i,j)$ for which $n_{ij}>0$. Associated with each edge $(i,j)$ is the random sample $\mathcal{Y}_{ij} = (Y_{ij1},\ldots,Y_{ijn_{ij}})$ of $n_{ij}$ comparisons, and let $\mathcal{Y}$ denote the collection of all such samples. The pair $(\mathcal{G},\mathcal{Y})$ is called a pairwise comparison graph (PCG).

In cardinal PCD, $Y_{ijk}=-Y_{jik}$ and $\nu_{ij}=-\nu_{ji}$ (Jiang et al., 2011). Thus, model \eqref{model.Y_ijk} is indexed by a parameter $\bnu \in \mathcal{N} \equiv \mathbb{R}^{K(K-1)/2}$, where \(\bnu = (\nu_{12},\ldots,\nu_{1K},\nu_{23},\ldots,\nu_{2K}, \ldots,\nu_{K-1,K}) \) is the vector of means arranged in lexicographic order. We refer to $\bnu$ as the preference profile. We write $i \succeq j$ when $\nu_{ij} \geq 0$, and $i \succ j$ when $\nu_{ij} > 0$. A preference relation is said to be transitive if $i \succeq j$ and $j \succeq k$ imply $i \succeq k$. When transitivity holds for all triplets, a weak global ranking exists. Various notions of stochastic transitivity have been studied in the context of paired comparison data, particularly for binary outcomes (see Fishburn, 1973; Oliveira et al., 2018). In particular, weak, strong, and linear stochastic transitivity (WST, SST, and LST) impose increasingly restrictive coherence conditions on triplets of items. A common modeling assumption is that, for all $1\leq i<j\leq K$,
\begin{equation} \label{nu.ij=mu.i-mu.j}
\nu_{ij}=\mu _{i}-\mu _{j},
\end{equation}
where $\mu_1,\ldots,\mu_K$ are referred to as scores or merits. In the cardinal setting, \eqref{nu.ij=mu.i-mu.j} is the linear analogue of LST and implies transitivity. Under \eqref{nu.ij=mu.i-mu.j}, \(
\bnu^{\top} = (\mu_1-\mu_2,\ldots,\mu_1-\mu_K,\mu_2-\mu_3,\ldots,\mu_2-\mu_K,\ldots,\mu_{K-1}-\mu_K)
\) for some $\bmu\in\mathbb{R}^{K}$. The collection of vectors satisfying \eqref{nu.ij=mu.i-mu.j} is denoted by $\mathcal{L}$. It is well known that the parameterization \eqref{nu.ij=mu.i-mu.j} by $\bmu$ is non--identifiable. Consequently, it is customary to impose an additional linear constraint such as \(\mu_{1}+\cdots+\mu_{K}=0\). Under this constraint, $\bmu$ is identifiable and $\mathcal{L}$ can be identified with $\mathbb{R}^{K-1}$. Model \eqref{nu.ij=mu.i-mu.j} and its least squares estimation have been studied extensively in statistics, operations research, and related fields (see, for example, Mosteller, 1951; Kwiesielewicz, 1996; Csató, 2015). A recent review is given by Singh et al. (2025). Departures from \eqref{nu.ij=mu.i-mu.j} may arise when the merits vary over time (dynamic models; Glickman, 2001; Baker and McHale, 2014) or depend on covariates (Dong et al., 2026). However, such effects do not by themselves constitute structural cyclicality in a fixed preference profile and are therefore outside the scope of the present work.

To date, a principled treatment of cyclicality and intransitivity in paired comparison models for ranking, prediction, and decision support remains limited. Existing work has largely developed along two complementary but separate directions. On the one hand, several authors have studied the structural properties of paired comparison data. Jiang et al. (2011) view cardinal PCGs as edge flows and, using discrete Hodge theory (Lim, 2020), decompose such flows into orthogonal components corresponding to gradient, curl, and harmonic flows. Saari (2014, 2021) develops an equivalent linear algebraic formulation and studies decompositions into linear and cyclic components. These works provide important structural and conceptual insights into cyclicality and intransitivity, but do not develop inferential procedures for estimating, testing, or recovering cyclic structure from observed data. On the other hand, the statistical literature has focused primarily on detecting or accommodating departures from transitivity. Early contributions include Kendall and Smith (1940), who introduced a coefficient of consistency based on combinatorial arguments, together with subsequent developments by Moran (1950), Alway (1962), Knezek et al. (1998), and Iida (2009). More recently, Singh and Davidov (2026) developed lack--of--fit tests for model \eqref{model.Y_ijk} under the assumption \eqref{nu.ij=mu.i-mu.j}. These methods are primarily diagnostic in nature and do not provide a framework for modeling or recovering the underlying cyclic structure. Other recent work has proposed parametric models allowing for forms of intransitivity, including Causeur and Husson (2005), Chen and Joachims (2016), Makhijani and Ugander (2019), Spearing et al. (2023), Zhang and Chen (2025), Okahara et al. (2026), and Lee and Chen (2026). While these approaches accommodate certain forms of intransitivity, they do not address the structural representation or recovery of minimal cyclic structure.

The present work bridges this gap by combining structural representations of paired comparison profiles with inferential and model selection methodology. Our main contributions are as follows:

\begin{enumerate}

\item We analyze the parameter space $\mathcal{N}$, deriving an orthogonal decomposition into transitive and cyclic components. The cyclic component is represented using an overcomplete dictionary of elementary cycles, leading to a geometric framework for minimal cyclic representations.

\item We formulate the recovery of cyclicality as a sparse model selection problem. The proposed tick--table representation motivates forward tick--based selection procedures for identifying minimal cyclic structure.

\item We establish large--sample guarantees for estimation and model recovery. The resulting procedures yield identifiable representations and consistently recover the underlying cyclic structure under suitable conditions.

\end{enumerate}

The paper is organized as follows. Section \ref{section:nu:decomposition} studies the decomposition of the parameter space $\mathcal{N}$. Inference and model selection are developed in Section \ref{select:section}. Section \ref{section:simulation} presents simulation results, and Section \ref{section:real:example} contains an illustrative example. We conclude in Section \ref{discussion} with a summary and discussion. Proofs are given in the Supplement. The \textsf{R} scripts used to reproduce the experiments are available at \href{https://github.com/rahulstats/GLM-II}{\texttt{github.com/rahulstats/GLM-II}}.

%%%%%%%%%%%%%%%%%%%%%%%%%%%%
%%%%%%%%%%%%%%%%%%%%%%%%%%%%
%%%%%%%%%%%%%%%%%%%%%%%%%%%%

\section{Cyclicality, intransitivity and minimal models} \label{section:nu:decomposition}

This section studies the structure of the parameter space $\mathcal{N}$ using an orthogonal decomposition into linear and cyclic components. Explicit spanning sets are derived for the corresponding subspaces, yielding explicit representations of cyclicality and intransitivity in paired comparison data. These structural results motivate the inferential and model selection procedures developed later in the paper. To present these ideas transparently, we assume throughout this section that the comparison graph $\mathcal{G}$ is complete and, consequently, that the full preference profile, i.e., the entire vector $\bnu$, is available. This assumption can be relaxed, as discussed in Section~\ref{discussion}.

%%%%%%%%%%%%%%%%%%%%%%%%%%%%

\subsection{Linear and cyclic decomposition}

Let $\bB$ be a $\binom{K}{2} \times K$ matrix with columns $\pmb{b}_1,\ldots,\pmb{b}_K$. The $(s,t)^{th}$ lexicographically ordered element of $\pmb{b}_k$, denoted $b_{k}(s,t)$, is  defined by the relation
\begin{align} \label{def:bij(k)}
b_{k}(s,t)=\mathbb{I}(s=k) -\mathbb{I}(t=k).
\end{align}
\noindent It is easily verified that if  $\bnu\in \mathcal{L}$ it can be expressed as
\begin{align} \label{nu:inM}
\bnu= (\mu_1-\mu_2,\ldots,\mu_{K-1}-\mu_K)^{\top}= \mu_1\pmb{b}_1+ \ldots +\mu_K\pmb{b}_K = \bB\bmu,
\end{align}
and therefore $\mathcal{L}=\mathrm{span}(\pmb{b}_1,\ldots,\pmb{b}_K)$. Consequently, $\mathcal{L}$ is a linear subspace of $\mathcal{N}$ for which $\{\pmb{b}_1,\ldots,\pmb{b}_K\}$ is a natural spanning set, see Saari (2021). It is also easy to verify that $\sum_{k=1}^{K}\pmb{b}_{k} = \pmb{0}$ and that the dimension of $\mathcal{L}$ is $K-1$. 

Jiang et al. (2011), using graphical combinatorial Hodge theory (see, Lim, 2020), and  Saari (2014) and Saari (2021), using the language of linear algebra, established that: 

\begin{fact} \label{linear:nu:restriction}
If
\begin{align} \label{eq:linear:nu:restriction}
\nu_{ij}+\nu_{jk}+\nu_{ki}=0
\end{align}
for all $(i,j,k)\in\{1,2,\ldots,K\}^3$ then $\bnu\in\mathcal{L}$.
\end{fact}

The preference vector $\bnu$ is said to be consistent on the triad $(i,j,k)$ if \eqref{eq:linear:nu:restriction} holds, and globally consistent if \eqref{eq:linear:nu:restriction} holds on every triad. Global consistency is equivalent to LST, i.e., the parametrization  \eqref{nu.ij=mu.i-mu.j}. 

If \eqref{eq:linear:nu:restriction} does not hold, then a cyclical preference relation between the items $i,j$ and $k$ is obtained. For example, if $(\nu_{ij},\nu_{ik},\nu_{jk})=(1,-1,1)$ then \eqref{eq:linear:nu:restriction} is obviously violated, moreover $i \succ j \succ k\succ i$. This preference relation is captured by a cyclical component of $\mathcal{N}$ that is outside of $\mathcal{L}$. Next, denote the orthogonal complement of $\mathcal{L}$ in $\mathcal{N}$ by $\mathcal{C}$. It follows that any $\bnu \in \mathcal{N}$ can be uniquely expressed as a sum of two orthogonal vectors, $\bnu_{\rm linear}\in \mathcal{L}$ and $\bnu_{\rm cyclic}\in \mathcal{C}$ for which 
\begin{align} \label{decomposition:nu}
\bnu = \bnu_{\rm linear} + \bnu_{\rm cyclic}
\end{align}
where $\bnu_{\rm linear}$ and $\bnu_{\rm cyclic}$ capture, respectively, the linear and cyclic aspects of the preference relations (Saari, 2021). 

\begin{example} \label{example:cyclic:relation}
Suppose that $K=3$ and $\bnu=(1,-1,1)$. Clearly $\nu_{12}+\nu_{23}+\nu_{31}=3$ so $\bnu \notin \mathcal{L}$. By \eqref{def:bij(k)}, $\pmb{b}_1 = (1,1,0)$, $\pmb{b}_2 = (-1,0,1)$ and $ \pmb{b}_3 = (0,-1,-1)$. Clearly $\bnu^{\top}\pmb{b}_i=0$ for all $i$ so $\bnu\in \mathcal{C}$ and consequently $\bnu_{\rm linear}=\bzero$ and $\bnu_{\rm cyclic}=\bnu$. 
Moreover, \(\nu_{12}=1,\quad \nu_{23}=1,\quad \nu_{13}=-1\), implying that
\(1\succ 2\succ 3\succ 1\). Thus, the preference relation is cyclical and non--transitive. A cyclic relation involving three items is called a cyclic triad.
\end{example}

The next example illuminates the relationship between cyclicality and various forms of transitivity. 

\begin{example} \label{nu:linear:cyclic:example}
Let $\bnu_{\rm linear}= (1,2,1)\in \mathcal{L}$ and $ \bnu_{\rm cyclic} = (1,-1,1)\in \mathcal{C}$. Clearly $\bnu_{\rm linear}$ is associated with the preference relation $1\succ 2\succ 3$ whereas $\bnu_{\rm cyclic}$ corresponds to the cyclic relation $1 \succ 2 \succ 3 \succ 1$, as demonstrated in Example \ref{example:cyclic:relation}. Next, let
\begin{align*}
\bnu = \alpha \bnu_{\rm linear} + \beta \bnu_{\rm cyclic}
\end{align*}
where $\alpha,\beta \ge 0$. A bit of algebra shows that \eqref{model.Y_ijk} is linearly transitive (or LST) if and only if $\beta=0$, strongly transitive (or SST) if $\alpha > 2\beta$ and weakly transitive (or WST) if $\alpha > \beta/2$. Moreover, if $\alpha = \beta/2$ then $1\succ 2$, $2\succ 3$ but $1 \sim 3$, i.e., $\nu_{13}=0$, whereas if $\alpha < \beta/2$ then the cyclical component dominates the linear component, i.e., $1\succ 2\succ 3\succ 1$. The model becomes purely cyclic if and only if $\alpha=0$.
\end{example}

\begin{remark}
Example \ref{nu:linear:cyclic:example} shows that cyclicality and intransitivity are closely related but conceptually distinct. Cyclicality refers to structural properties of the preference profile encoded through the cyclic components, whereas intransitivity refers to the induced behavior of the preference relation itself. In particular, a preference relation may have cyclic components yet be transitive.     
\end{remark}

Example \ref{example:cyclic:relation} shows that when $K=3$ the cyclic relation $1\succ 2\succ 3\succ 1$ can be expressed by the vector $(1,-1,1)$ which is orthogonal to $\mathcal{L}$ and completes the basis for $\mathcal{N}$. Similarly, for any $K$ the cyclic relation $i\succeq j\succeq k\succeq i$ can be expressed by the vector $\pmb{c}_{(i,j,k)} \in \mathcal{C}$ whose $(s,t)^{th}$ element, in lexicographical order, is
\begin{align} \label{cyclic:triangle:not:unique}
c_{(i,j,k)}(s,t) = \mathbb{I}((s,t)\in\{(i,j),(j,k),(k,i)\}) - \mathbb{I}((s,t)\in\{(j,i),(k,j),(i,k)\}). 
\end{align}
Thus, the $(i,j)^{th}$ and $(j,k)^{th}$ elements of $\pmb{c}_{(i,j,k)}$ are $1$; the $(i,k)^{th}$ element is $-1$, and all other elements are $0$. Hence, cyclical preference relations admit a linear algebraic representation generated by elementary cyclic triads. It is easy to verify that
\[
\pmb{c}_{(i,j,k)} = \pmb{c}_{(j,k,i)} = \pmb{c}_{(k,i,j)} = -\pmb{c}_{(j,i,k)} = -\pmb{c}_{(i,k,j)} = -\pmb{c}_{(k,j,i)},
\]
so cyclic relations differing only by orientation span the same one--dimensional subspace. Consequently, it suffices to consider only vectors $\pmb{c}_{(i,j,k)}$ with $i<j<k$. Further note that a cyclic relation on $(i,j,k)$ can be expressed as a superposition of the cyclic relations $(s,i,j),(s,j,k)$ and $(s,i,k)$ for some $s \notin \{i,j,k\}$, i.e., 
\begin{align}\label{cijk:relation:with:other:c}
\pmb{c}_{(i,j,k)}= \pmb{c}_{(s,i,j)}+ \pmb{c}_{(s,j,k)}- \pmb{c}_{(s,i,k)}, 
\end{align}
showing that cyclic triads are linearly dependent whenever $K\ge4$. 
The following was established by Saari (2014) and Saari (2021),
\begin{fact} \label{span:M-N:set:fact}
The set $\{\pmb{c}_{(i,j,k)}\}_{1\leq i<j<k\leq K}$, whose cardinality is $\binom{K}{3}$, spans $\mathcal{C}$. Furthermore,  
$$ \dim(\mathcal{C})=\binom{K}{2}-(K-1)=\binom{K-1}{2}.$$  
\end{fact}

Let $\bC$ denote the $\binom{K}{2}\times \binom{K}{3}$ matrix whose columns are the vectors $\{\pmb{c}_{(i,j,k)}\}_{1\leq i<j<k\leq K}$. The rows and columns of $\bC$ are indexed by edges and triads, respectively. Following Saari (1989), the columns of $\bC$ may be viewed as a dictionary of elementary cyclical relations. It follows that for $\bnu\in\mathcal{C}$ we have
\begin{align} \label{nu:NminusM}
\bnu = \sum_{1\leq i<j<k\leq K}\gamma_{ijk}\,\pmb{c}_{(i,j,k)} = \bC\pmb{\gamma},
\end{align}
where $\gamma_{ijk}$ is the coefficient associated with the cyclical triad $(i,j,k)$ and $\pmb{\gamma}\in\mathbb{R}^{\binom{K}{3}}$. Since $\dim(\mathcal{C}) < \binom{K}{3}$ whenever $K\ge4$, representations of $\bnu$ in terms of cyclic triads need not be unique. This naturally leads to the problem of identifying parsimonious cyclic representations.

%%%%%%%%%%%%%%%%%%%%%%%%%%%%

\subsection{Minimal models and tick--tables}

Since $\mathcal{L} \perp\!\!\!\perp \mathcal{C}$, the linear and cyclical components of $\bnu\in\mathcal{N}$ are given by
\begin{align}\label{components:of:bnu}
\bnu_{\rm linear}= \bB (\bB^{\top} \bB)^{+} \bB^{\top} \bnu
\quad \text{and} \quad
\bnu_{\rm cyclic}= \bC (\bC^{\top}\bC)^{+}\bC^{\top}\bnu.
\end{align}
Thus, every preference profile admits a unique decomposition into linear and cyclical components. Equivalently, every $\bnu\in\mathcal{N}$ admits the parametrization
\begin{align} \label{bnu:decomposed:linear:cyclic}
\bnu
=
\bnu_{\rm linear}
+
\bnu_{\rm cyclic}
=
\bB\bmu+\bC\pmb{\gamma},
\end{align}
where $\bmu$ parametrizes the linear component and $\pmb{\gamma}$ parametrizes the cyclical component. The parameter $\bmu$ is identifiable once a constraint such as $\sum_{i=1}^K\mu_i=0$ is imposed; see Singh et al. (2025). In contrast, the cyclic representation $\pmb{\gamma}$ is generally not identifiable. Indeed, Equation \eqref{cijk:relation:with:other:c} and Fact \ref{span:M-N:set:fact} imply that the set
\[
\Gamma_{\bnu_{\rm cyclic}} = \{\pmb{\gamma}\in\mathbb{R}^{\binom{K}{3}}:
\bnu_{\rm cyclic}=\bC\pmb{\gamma}\}
\]
contains multiple elements.

\begin{definition} \label{def:minimal:model}
For $\pmb{\gamma}\in \Gamma_{\bnu_{\rm cyclic}}$ let $s(\pmb{\gamma}) = \sum\mathbb{I}(\gamma_{ijk}\neq 0)$ denote the size of the model associated with $\pmb{\gamma}$. A model $\pmb{\gamma}\in\Gamma_{\bnu_{\rm cyclic}}$ is said to be minimal if $s(\pmb{\gamma})\leq s(\pmb{\gamma}')$ for all $\pmb{\gamma}'\in \Gamma_{\bnu_{\rm cyclic}}$.
\end{definition}

To illustrate the concept of a minimal model, suppose that $K=4$ and $\bnu_{\rm cyclic}=\pmb{c}_{(2,3,4)}$. Then $\pmb{\gamma}_1=(0,0,0,1)\in\Gamma_{\bnu_{\rm cyclic}}$. Moreover, by \eqref{cijk:relation:with:other:c}, $\bnu_{\rm cyclic}=\pmb{c}_{(1,2,3)}+\pmb{c}_{(1,3,4)}-\pmb{c}_{(1,2,4)}$, so that $\pmb{\gamma}_2=(1,1,-1,0)\in\Gamma_{\bnu_{\rm cyclic}}$. Since $s(\pmb{\gamma}_1)=1$ and $s(\pmb{\gamma}_2)=3$, the representation $\pmb{\gamma}_1$ is minimal by Definition \ref{def:minimal:model}. Thus, identifying a minimal model amounts to finding the sparsest, or equivalently, the most parsimonious representation of $\bnu_{\rm cyclic}$ in the dictionary generated by the cyclical triads.  This requires considering the full collection $\{\pmb{c}_{(i,j,k)}\}_{1\leq i<j<k\leq K}$ of possible cyclic triads. Throughout the paper, we assume that the data--generating mechanism is associated with a minimal model. Although this assumption is generally not directly testable, it provides a natural notion of structural parsimony. Associated with each minimal model is the collection $\mathcal{C}_{\pmb{\gamma}}$ of columns of $\bC$ corresponding to the nonzero coefficients of $\pmb{\gamma}$.

\begin{definition}
The span and support of a minimal model are
$${\rm span}(\mathcal{C}_{\pmb{\gamma}})= {\rm span}(\{\pmb{c}_{(i,j,k)}:\, \gamma_{ijk}\neq 0 \}) ~~\mbox{and}~~  
{\rm support}(\mathcal{C}_{\pmb{\gamma}})= \{(i,j):  \pmb{c}_{(i,j,k)}\in {\rm span}(\mathcal{C}_{\pmb{\gamma}}) \}.$$ 
\end{definition}  

The support describes the edges involved in the cyclic structure, whereas the span characterizes its underlying linear representation. Ideally, minimal models would have unique support and be identifiable from their span.

\begin{example} \label{eg:multiple:gamma}
For any $K\geq 6$ let $\bnu_{\rm cyclic}= \pmb{c}_{(1,2,3)}-\pmb{c}_{(1,2,4)}+ \pmb{c}_{(1,3,6)}- \pmb{c}_{(2,3,5)} $. We refer to this model as $\mathcal{C}_{\pmb{\gamma}_1}$. Using \eqref{cijk:relation:with:other:c} we also express $\bnu_{\rm cyclic}$ as $-\pmb{c}_{(1,3,4)}+ \pmb{c}_{(1,3,6)}+ \pmb{c}_{(2,3,4)}- \pmb{c}_{(2,3,5)}$ which we refer to as $\mathcal{C}_{\pmb{\gamma}_2}$. Thus, the minimal model is not unique. Moreover, ${\rm span}(\mathcal{C}_{\pmb{\gamma}_1}) \neq{\rm span}(\mathcal{C}_{\pmb{\gamma}_2})$ and ${\rm support}(\mathcal{C}_{\pmb{\gamma}_1}) \neq{\rm support}(\mathcal{C}_{\pmb{\gamma}_2})$ so minimal models may be associated with different linear spans and supports.
\end{example}

Example \ref{eg:multiple:gamma} shows that minimal models need not be unique. Non--uniqueness arises from the special algebraic structure of overlapping cyclic triads together with exact cancellations among the corresponding coefficients. Consequently, non--unique minimal models are unlikely to occur in practice.

\begin{theorem} \label{thm:unique:minimal:model}
For minimal models $\pmb{\gamma}_1$ and $\pmb{\gamma}_2$ we have $\bnu_{\rm cyclic}\in {\rm span}(\mathcal{C}_{\pmb{\gamma}_1}) \cap {\rm span}(\mathcal{C}_{\pmb{\gamma}_2})$. Moreover, if ${\rm span}(\mathcal{C}_{\pmb{\gamma}_1}) = {\rm span}(\mathcal{C}_{\pmb{\gamma}_2})$ then $ {\rm support}(\mathcal{C}_{\pmb{\gamma}_1})={\rm support}(\mathcal{C}_{\pmb{\gamma}_2})$. 
Furthermore, suppose the following conditions are satisfied:
\begin{itemize}
  \item[(i)] For all $(i, j) \in \operatorname{supp}(\mathcal{C}_{\pmb{\gamma}})$, we have $\nu_{\text{cyclic}, ij} \neq 0$;
  \item[(ii)] If $\operatorname{supp}(\mathcal{C}_{\pmb{\gamma}_1}) = \operatorname{supp}(\mathcal{C}_{\pmb{\gamma}_2})$, then $\operatorname{span}(\mathcal{C}_{\pmb{\gamma}_1}) = \operatorname{span}(\mathcal{C}_{\pmb{\gamma}_2})$,
\end{itemize}
then $\mathcal{C}_{\pmb{\gamma}_1}$ and $\mathcal{C}_{\pmb{\gamma}_2}$ span the same subspace. Finally, if any two triads in a minimal model share at most one index then the minimal model is unique.
\end{theorem}

Theorem \ref{thm:unique:minimal:model} shows that minimal models are often identifiable from their support. Moreover, even when minimal models are not unique, their spans frequently coincide.

Note that $\nu_{{\rm cyclic},ij}$, the $(i,j)^{th}$ component of $\bnu_{\rm cyclic}$, is given by
\begin{equation}\label{nu:cyclic:ij}
\nu_{{\rm cyclic},ij}
=
\sum_{1\leq s<t<u\leq K}
c_{(s,t,u),ij}\gamma_{stu},
\end{equation}
where $c_{(s,t,u),ij}$ is the $(i,j)^{th}$ component of $\pmb{c}_{(s,t,u)}$. In particular, $(i,j)\in {\rm support}(\mathcal{C}_{\pmb{\gamma}})$ implies that the contributions of the active cyclic triads do not cancel at the edge $(i,j)$. Exact cancellations require highly structured coefficient configurations and are therefore unlikely to occur in practice.

\medskip

It can be easily verified that for any triad $(i,j,k)$ 
\begin{align*}
\nu_{ij} +\nu_{jk} +\nu_{ki} = \nu_{{\rm cyclic},ij}+ \nu_{{\rm cyclic},jk}+ \nu_{{\rm cyclic},ki}. 
\end{align*}

Thus, cyclic triads, i.e., triads violating \eqref{eq:linear:nu:restriction}, can also be identified using $\bnu_{\rm cyclic}$. However, not every cyclic triad belongs to a minimal model. For example, if $\bnu_{\rm cyclic}=\pmb{c}_{(1,2,3)}$, then any triad $(i,j,k)$ sharing two indices with the triad $(1,2,3)$, e.g., $(1,2,4)$, will be a cyclic triad, but not be part of any minimal model. 

\begin{definition}
Let $\mathcal{T}_{\bnu_{\rm cyclic}}$ denote a table whose rows and columns are indexed by edges and triads, respectively. A tick mark is placed in position $((s,t),(i,j,k))$ whenever $\nu_{{\rm cyclic},st}\neq 0$ and the edge $(s,t)$ belongs to the triad $(i,j,k)$. The resulting table is referred to as the tick--table associated with $\bnu_{\rm cyclic}$.
\end{definition}

The columns of $\mathcal{T}_{\bnu_{\rm cyclic}}$ may contain $0$, $1$, $2$, or $3$ ticks. For example, let $K=5$ and set $ \bnu_{\rm cyclic} = \pmb{c}_{(1,2,3)}$, then $\mathcal{T}_{\bnu_{\rm cyclic}}$ is given by: 

\FloatBarrier
\begin{table}[ht]
\centering
\begin{tabular}{rllllllllll}
 & 123 & 124 & 125 & 134 & 135 & 145 & 234 & 235 & 245 & 345 \\ 
  12 & \checkmark & \checkmark & \checkmark &  &  &  &  &  &  &  \\ 
  13 & \checkmark &  &  & \checkmark & \checkmark &  &  &  &  &  \\ 
  14 &  &  &  &  &  &  &  &  &  &  \\  
  15 &  &  &  &  &  &  &  &  &  &  \\ 
  23 & \checkmark &  &  &  &  &  & \checkmark & \checkmark &  &  \\ 
  24 &  &  &  &  &  &  &  &  &  &  \\ 
  25 &  &  &  &  &  &  &  &  &  &  \\ 
  34 &  &  &  &  &  &  &  &  &  &  \\ 
  35 &  &  &  &  &  &  &  &  &  &  \\ 
  45 &  &  &  &  &  &  &  &  &  &  \\
\end{tabular}
\end{table}
\FloatBarrier

\noindent Note that in this tick--table only the $(1,2,3)^{th}$ column, corresponding to the minimal model, contains three ticks, whereas all other columns contain one or zero ticks.

\begin{theorem} \label{thm:ticks:real:model}
If there is a minimal model $\pmb{\gamma}$ with some $0$--tick triads then there exists another minimal model $\pmb{\gamma}'$ for which no triad has $0$-ticks. Additionally, if any two triads in a minimal model share at most one index then only triads in the minimal model will have $3$--ticks. However not all $3$--tick triads are members of a minimal model. Finally, if the minimal model is unique and some of its triads share two indices then some of the triads may have less than $3$--ticks.
\end{theorem}

Theorem \ref{thm:ticks:real:model} shows that the tick--table encodes information about the underlying minimal cyclic structure. In particular, when the minimal model is unique, the tick--table may be used to identify the generating triads. More generally, it may be used to recover at least one admissible minimal representation. Consequently, the tick--table provides a combinatorial description of the underlying cyclic structure and forms the basis for the recovery procedures developed later in the paper.

%%%%%%%%%%%%%%%%%%%%%%%%%%%%
%%%%%%%%%%%%%%%%%%%%%%%%%%%%
%%%%%%%%%%%%%%%%%%%%%%%%%%%%

\section{Model selection} \label{select:section}

The decomposition of $\mathcal{N}$ together with the notions of minimal models and tick--tables introduced in Section \ref{section:nu:decomposition} provide a framework for inferring cyclic structure from paired comparison data. To this end, we develop methods for estimation and evaluation of candidate models and introduce forward tick--based selection (FTBS), a procedure for recovering sparse cyclic structure from paired comparison data.

%%%%%%%%%%%%%%%%%%%%%%%%%%%%

\subsection{Intermediate models: estimation and tests for lack of fit}

If $\bnu \in \mathcal{L}$, we say that the reduced model holds. An intermediate model corresponds to $\bnu \in \mathcal{S}$ where $\mathcal{L}\subsetneq \mathcal{S}\subsetneq \mathcal{N}$ and $\subsetneq$ denotes strict inclusion. In this case, $\mathcal{S} = {\rm span}(\bB,\bC_s)$ where $\bC_s$ is a sub--matrix of $\bC$ with at most $\binom{K-1}{2}-1$ columns. Otherwise, the full model holds.

Let $\bY$ be the $n \times 1$ vector of outcomes arranged lexicographically, i.e.,  
\begin{align*}
\bY = (Y_{121},\ldots,Y_{12n_{12}}, Y_{131},\ldots,Y_{13n_{13}}, \ldots, Y_{K-1,K1},\ldots,Y_{K-1,Kn_{K-1,K}})^{\top}.
\end{align*}
Let $\bL$ denote the $n\times|\bnu|$ incidence matrix mapping observations to the corresponding components of $\bnu$, i.e., $L_{st}=1$ if the $s^{th}$ observation corresponds to the $t^{th}$ component of $\bnu$, and $L_{st}=0$ otherwise, in which case $\mathbb{E}(\bY)=\bL\bnu$. Consequently, every intermediate model can be expressed as 
\begin{align} \label{joint:nu:model:simplified}
\bY = (\bL\bB,\,\bL\bC_{s})\begin{pmatrix}
    \bmu\\ \pmb\gamma_{s}
\end{pmatrix} + \pmb{\epsilon},
\end{align}
where $\bC_s$ is a minimum--column sub--matrix of $\bC$ spanning $\mathcal{S}$ and $\pmb{\gamma}_s$ is the corresponding coefficient vector. The LSE is defined by
\[
(\widehat{\bmu},\widehat{\pmb{\gamma}}_s)
=
\arg\min_{\bmu,\pmb{\gamma}_s}
\{ Q(\bmu,\pmb{\gamma}_s): \b1^{\top}\bmu=0 \},
\]
where $Q(\bmu,\pmb{\gamma}_s)= (\bY-\bnu(\bmu,\pmb{\gamma}_s))^{\top}
(\bY-\bnu(\bmu,\pmb{\gamma}_s))$ is the sum of squares associated with \eqref{joint:nu:model:simplified}. Standard least squares calculations show that
\begin{align}
\label{mu:thm4.1}
\widehat\bmu
&=
\bN^{+}\bB^{\top}\bL^{\top}
(\bY- \bL\bC_s\widehat{\pmb\gamma}_s), \\
\label{nu:thm4.1}
\widehat{\pmb\gamma}_s
&=
( \bC_s^{\top}\bL^{\top}
(\pmb{I}-\bL\bB \bN^{+}\bB^{\top}\bL^{\top})
\bL\bC_s)^{+}
\bC_s^{\top}\bL^{\top}
(\pmb{I}- \bL\bB \bN^{+}\bB^{\top}\bL^{\top})
\bY,
\end{align}
where $\bN^{+}$ is the Moore--Penrose inverse of $\bN$, the Laplacian of the graph $\mathcal{G}$. When $\bC_s=\pmb{0}$, the estimator \eqref{mu:thm4.1} reduces to $\bN^{+}\bS$, where $\bS=(S_{1},\ldots,S_{K})^{\top}$ with components $S_{i}=\sum_{j\neq i}S_{ij}$ and $S_{ij}=\sum_{k=1}^{n_{ij}}Y_{ijk}$. We will assume that $n_{ij}>0$ for all $1\leq i<j\leq K$, and that $n_{ij}\to\infty$ with $n_{ij}=O(n)$. 

\begin{theorem} \label{thm:mu:lambda:estimation:properties}
Assume that the errors $\epsilon_{ijk}$ are IID with zero mean and finite variance $\sigma^2$ and suppose the minimal model is in $\mathcal{S}$. If so the LSEs in \eqref{mu:thm4.1} and \eqref{nu:thm4.1} are unbiased and
\[ 
\sqrt{n}\begin{pmatrix}
\widehat{\bmu}-\bmu\\ \widehat{\pmb\gamma}_s- \pmb{\gamma}_s
\end{pmatrix} \Rightarrow \mathcal{N}(\bzero, \pmb{\Sigma}_{\mu,\gamma_s}),\]
with
\begin{align*}
\pmb{\Sigma}_{\mu,\gamma_s} = \sigma^2 \begin{pmatrix}
\bB^{\top}\pmb{\Xi} \bB & \bB^{\top}\pmb{\Xi} \bC_s \\
\bC_s ^{\top}\pmb{\Xi} \bB & \bC_s ^{\top}\pmb{\Xi} \bC_s 
\end{pmatrix}^{+} \begin{pmatrix}
\bB^{\top}\\ \bC_s ^{\top}
\end{pmatrix}\pmb{\Xi}^3 
(\bB, \bC_s) \begin{pmatrix}
\bB^{\top}\pmb{\Xi} \bB & \bB^{\top}\pmb{\Xi} \bC_s \\
\bC_s ^{\top}\pmb{\Xi} \bB & \bC_s ^{\top}\pmb{\Xi} \bC_s 
\end{pmatrix}^{+}.
\end{align*}
where $\pmb{\Xi}$ is the $|\bnu| \times |\bnu|$ diagonal matrix whose $(i,j)^{th}$ diagonal element is $\theta_{ij}=\lim {n_{ij}/n}$.
\end{theorem}

Theorem \ref{thm:mu:lambda:estimation:properties} shows that when the minimal model is contained in $\mathcal{S}$, the LSE $(\widehat{\bmu},\widehat{\pmb{\gamma}}_s)$ is unbiased, consistent, and asymptotically normal. Recovering cyclic structure, however, requires methods for determining whether a proposed intermediate model adequately explains the observed preference relations. This motivates the study of lack--of--fit testing for intermediate cyclic models, which underlies the sequential evaluation of candidate models within FTBS. Consider testing $H_0:\bnu\in\mathcal{S}$ against $H_1:\bnu\notin\mathcal{S}$. Motivated by Jiang et al. (2011), see also Singh and Davidov (2026), define
\begin{align*}
R_{n,\mathcal{S}}
=
\pmb{U}_{n,\mathcal{S}}^{\top}\pmb{U}_{n,\mathcal{S}},
\end{align*}
where $\pmb{U}_{n,\mathcal{S}}=\bD^{1/2}(\widehat{\bnu}-\widehat{\bnu}_s)$. Here $\bD$ is the $|\bnu|\times|\bnu|$ diagonal matrix whose $(i,j)^{th}$ diagonal element is $n_{ij}$; and $\widehat{\bnu}$ is the vector whose $(i,j)^{th}$ element is $S_{ij}/n_{ij}$, and $\widehat{\bnu}_s=\bB\widehat{\bmu}+\bC_s\widehat{\pmb{\gamma}}_s$ is the LSE of $\bnu$ under $H_0$. Thus, $R_{n,\mathcal{S}}$ is the squared norm of the residual vector.

\begin{theorem} \label{Thm-intermediate-model-gof}
Assume that $\bnu\in \mathcal{S}$ and the condition of Theorem \ref{thm:mu:lambda:estimation:properties} hold. If so as $n \to \infty$
\begin{equation*}
R_{n,\mathcal{S}}\Rightarrow \sum_{i=1}^{t}{\lambda}_{i} Z_{i}^{2},
\end{equation*}
where $t={\rm rank}({\pmb{\Psi }}_{\mathcal{S}})=\binom{K}{2}-(K+r-1)$, $r$ is the number of columns in $\bC_s$, $Z_{1},\ldots ,Z_{t}$ are independent $\mathcal{N}(0,1)$ RVs and ${\lambda} _{1},\ldots ,{\lambda}_{t}$ are the non--zero eigenvalues of the $|\bnu|\times|\bnu|$ matrix
${\pmb{\Psi }}_{\mathcal{S}}= \sigma^2\, 
\pmb{\Xi}^{1/2} \pmb{M}\,\pmb{\Xi}^{+} \pmb{M}^{\top}\pmb{\Xi}^{1/2}$,
where $$\pmb{M}= \pmb{I} - (\bB,\bC_s)\begin{pmatrix}
\bB^{\top}\pmb{\Xi} \bB & \bB^{\top}\pmb{\Xi} \bC_s \\
\bC_s ^{\top}\pmb{\Xi} \bB & \bC_s ^{\top}\pmb{\Xi} \bC_s 
\end{pmatrix}^{+} \begin{pmatrix}
\bB^{\top}\\ \bC_s ^{\top}
\end{pmatrix}\pmb{\Xi}.$$
\end{theorem}
\noindent Under local alternatives, Proposition \ref{prop:dist:alt:Rn:Wn} in the Supplement shows that
\begin{equation*}
R_{n,\mathcal{S}} \Rightarrow \sum_{i=1}^{t}\lambda_i(Z_i+\gamma_i)^2,
\end{equation*}
where $\gamma_1,\ldots,\gamma_t$ are the components of $\pmb{O}_s(\pmb{\Psi}_s^{+})^{1/2}\pmb{\Xi}^{1/2}\pmb{\delta}$ corresponding to the nonzero eigenvalues of $\pmb{\Psi}_s$, and $\pmb{O}_s$ is the orthonormal matrix whose columns are the eigenvectors of $\pmb{\Psi}_s$. Consequently, the resulting tests are consistent.

%%%%%%%%%%%%%%%%%%%%%%%%%%%%

\subsection{Forward tick--based selection}

Model selection methods for linear models include best subset selection, forward and backward stepwise regression, and regularization procedures such as the LASSO (Tibshirani 1996; Hastie et al. 2009; Freijeiro‐-González et al. 2022). Such procedures are often combined with evaluation criteria including AIC, BIC, and the adjusted--$R^2$; see Chen et al. (2013) and Claeskens and Hjort (2008).

The preceding results motivate a sequential procedure for recovering sparse cyclic structure in PCGs. Specifically, the combinatorial structure encoded by the tick--table is used to generate candidate intermediate models, while the lack--of--fit tests developed above determine whether these models adequately explain the observed preference relations. This leads to forward tick--based selection (FTBS), a procedure built on the structural results of Section \ref{section:nu:decomposition}, specifically Theorems \ref{thm:unique:minimal:model} and \ref{thm:ticks:real:model}. Although FTBS may be viewed as a relative of forward stepwise regression, it relies on the estimated tick--table rather than on features of the design matrix. The procedure combines local edgewise inference with global lack--of--fit testing to recover intermediate cyclic models. To fix ideas, let $\mathcal{S}\subseteq\mathcal{N}$ denote the smallest subspace containing all possible minimal models and let $\widehat{\mathcal{S}}$ denote the selected model. To construct $\widehat{\mathcal{S}}$, first compute $\widehat{\bnu}_{\rm cyclic}=\textswab{C}\,\widehat{\bnu}$ where $\widehat{\bnu}$ was described earlier and $\textswab{C}= \bC (\bC^{\top}\bC)^{+}\bC^{\top}$, see \eqref{components:of:bnu}. 

Let $\mathcal{T}_{\bnu_{\rm cyclic}^*}$ denote the tick--table corresponding to $\bnu_{\rm cyclic}^*$ where 
\begin{equation}\label{phi_ij:nu_cyclic}
\nu_{{\rm cyclic},ij}^*= 
\begin{cases}
\widehat\nu_{{\rm cyclic},ij} &\text{if } \phi_{ij}(\widehat\bnu_{{\rm cyclic}})=1,\\
0 &\text{if } \phi_{ij}(\widehat\bnu_{{\rm cyclic}})=0.
\end{cases}
\end{equation}
Here $\phi_{ij}(\widehat\bnu_{{\rm cyclic}})=1$ if the hypothesis $H_0^{(i,j)}:\nu_{{\rm cyclic},ij}=0$ is rejected and $0$ otherwise. The family of hypotheses $\{H_0^{(i,j)}\}_{1\le i <j \le K}$ is tested at level $\alpha$ while controlling either the family--wise error rate (FWER) or the false discovery rate (FDR); see Stoica and Babu (2022) for additional discussion of testing--based approaches to model selection.

\begin{remark} \label{remark:4.1}
It is readily seen that
\[
\sqrt{n}(\widehat{\bnu}_{\rm cyclic}- \bnu_{\rm cyclic})
\Rightarrow
\mathcal{N}(\bzero, \sigma^2 \textswab{C}\,\pmb{\Xi}^{+}\textswab{C}^{\top}).
\]
Consequently, $H_0^{(i,j)}:\nu_{{\rm cyclic},ij}=0$ can be tested using the statistic $\widehat\nu_{{\rm cyclic},ij}/\widehat{\sigma}_{{\rm cyclic},ij}$, where $\widehat{\sigma}_{{\rm cyclic},ij}$ is the estimated standard error of $\widehat\nu_{{\rm cyclic},ij}$. Let $\alpha_{ij}^{*}$ denote the corresponding $p$--value. The simplest procedure for controlling the FWER is the Bonferroni correction, i.e., reject those hypotheses for which $\alpha_{ij}^{*}<\alpha/\binom{K}{2}$. Controlling the FDR at level $\alpha$ is accomplished by rejecting only those hypotheses for which $\alpha_{ij}^{*}< \alpha\, r_{ij}^*/\binom{K}{2}$, where $r_{ij}^*$ is the rank of $\alpha_{ij}^{*}$ among all $p$--values.
\end{remark}

Let $\mathcal{C}_0,\ldots,\mathcal{C}_3$ denote the collections of columns of $\mathcal{T}_{\bnu_{\rm cyclic}}$ containing $0,\ldots,3$ ticks respectively. Further, let $\mathcal{C}_{1,\rm lin}$, $\mathcal{C}_{2,\rm lin}$, and $\mathcal{C}_{3,\rm lin}$ denote linearly independent subsets of $\mathcal{C}_{1}$, $\mathcal{C}_{2}$, and $\mathcal{C}_{3}$ respectively. Next, let $\mathcal{C}_{2,\rm lin}^{+}$ denote the set of elements in $\mathcal{C}_{2,\rm lin}$ which are not spanned by $\mathcal{C}_{3,\rm lin}$ and similarly let $\mathcal{C}_{1,\rm lin}^{+}$ denote the set of elements in $\mathcal{C}_{1,\rm lin}$ which are not spanned by $\mathcal{C}_{3,\rm lin}\cup \mathcal{C}_{2,\rm lin}^{+}$. Define the nested sequence of models
\(\mathcal{S}_1\subset \cdots \subset \mathcal{S}_4\) where
\[
\mathcal{S}_1=\mathcal{L}, \qquad
\mathcal{S}_{2}={\rm span}(\mathcal{S}_1\cup \mathcal{C}_{3,\rm lin}), \qquad
\mathcal{S}_{3}={\rm span}(\mathcal{S}_2\cup \mathcal{C}_{2,\rm lin}^{+}), \qquad
\mathcal{S}_{4}={\rm span}(\mathcal{S}_3\cup \mathcal{C}_{1,\rm lin}^{+}).
\]
Let $\mathcal{S}_{1}^{*},\ldots,\mathcal{S}_{4}^{*}$ denote the corresponding random sets constructed from $\mathcal{T}_{\bnu_{\rm cyclic}^{*}}$. We are now ready to define FTBS.

\FloatBarrier
\begin{algorithm}[ht]
  \caption*{Forward Tickwise Selection ($\rm FTBS$)} 
  \begin{algorithmic}[1]
    \State Test whether $H_0: \mathcal{S}=\mathcal{S}_1^*$ holds using Theorem \ref{Thm-intermediate-model-gof}. If the null is not rejected, set $\widehat{\mathcal{S}}=\mathcal{S}_1^*$ and stop. Otherwise, set $a=2$.
    \State Test whether $H_0: \mathcal{S}=\mathcal{S}_a^*$ holds using Theorem \ref{Thm-intermediate-model-gof}. If the null is not rejected, set $\widehat{\mathcal{S}}=\mathcal{S}_a^*$ and stop. Otherwise,
    \State Set $a=a+1$ and repeat Step 2.
  \end{algorithmic}
\end{algorithm}
\FloatBarrier

Algorithm FTBS is computationally simple. It requires $\binom{K}{2}$ elementary tests, see \eqref{phi_ij:nu_cyclic}, followed by the construction of the linearly independent sets $\mathcal{S}_{1}^{*},\ldots,\mathcal{S}_{4}^{*}$, which is computationally straightforward. In addition, FTBS requires fitting and evaluating, through lack--of--fit testing, at most four linear models; see Theorem \ref{Thm-intermediate-model-gof}. The theoretical properties of FTBS are investigated in Theorem \ref{thm:select:triad:convergence} below.

\begin{theorem} \label{thm:select:triad:convergence}
Fix $\alpha$ and let $\widehat{\mathcal{S}}$ denote the model selected by Algorithm $\rm FTBS$. If $n_{ij}\to\infty$, ${1\leq i<j\leq K}$ and $n_{ij}=O(n)$ then $\mathbb{P}(\bnu_{\rm cyclic}\in{\rm span}(\widehat{\mathcal{S}}))\to 1$. Moreover, if the errors are sub--Gaussian RVs, then for some $C_1,C_2\geq 0$ we have
$$\mathbb{P}(\bnu_{\rm cyclic}\in {\rm span}(\widehat{\mathcal{S}}))\geq 1- C_1 e^{(-C_2n)}.$$ 
Additionally, if: $(i)$ all tests within $\rm FTBS$ are performed at a level $\alpha_n\rightarrow 0$; (ii) the power of rejecting any of the false nulls is $\pi_n \rightarrow 1$ when $(i)$ holds; and if $(iii)$ triads in the minimal model share at most index. Then,
\begin{align} \label{eq.alpha.pi}
\mathbb{P}(\mathcal{S}=\widehat{\mathcal{S}})\to 1.
\end{align}  
\end{theorem}

Theorem \ref{thm:select:triad:convergence} shows that FTBS selects a model $\widehat{\mathcal{S}}$ satisfying
$\mathbb{P}\{\bnu_{\rm cyclic}\in{\rm span}(\widehat{\mathcal{S}})\}\to1$.
Moreover, if the minimal model is unique and the tests within FTBS are performed at vanishing significance levels, then
$\mathbb{P}(\mathcal{S}=\widehat{\mathcal{S}})\to 1$.
Under sub--Gaussian errors, the probability of failing to recover a model spanning the true cyclic component decays exponentially fast.

We refer to this procedure as vanilla FTBS. In the numerical studies below we use the FDR procedure of Benjamini and Hochberg (1995); see also Zhao and Sun (2025). Since vanilla FTBS accepts entire tick--blocks simultaneously, it may include redundant triads in finite samples. To encourage parsimony, we also consider a pruned version of FTBS. After a screened candidate block passes the lack--of--fit test, a recursive backward elimination step removes conditionally insignificant triads one at a time. This pruning step is in the spirit of Zheng and Loh (1995). We use pruned FTBS in the numerical studies as a finite--sample refinement of vanilla FTBS. Finally, we consider a hybrid LASSO+FTBS procedure. LASSO provides a computationally efficient screening device and can perform well in small samples, but, as shown in Section \ref{section:simulation}, it may select too many triads. Conversely, pruned FTBS favors parsimonious models, but its ability to recover the correct support can be limited for small sample sizes. To combine the complementary strengths of the two methods, LASSO is first used to screen candidate cyclic components; the corresponding tick--table is then constructed and evaluated using vanilla FTBS.

%%%%%%%%%%%%%%%%%%%%%%%%%%%%

\section{Simulations} \label{section:simulation}

We evaluate the competing model selection procedures
using both structural and performance criteria. Structural accuracy is
measured by support recovery and parsimony. When the minimal model is unique, support recovery is quantified by $\widehat{\mathbb{P}}(\mathcal{S}\subseteq \widehat{\mathcal{S}})$. When the minimal representation is not unique, this criterion is replaced by $\widehat{\mathbb{P}}(\bnu_{\rm cyclic}\in {\rm span}(\widehat{\mathcal{S}}))$, which records whether the selected model spans the true cyclic component. Parsimony is assessed by the expected relative model size of the selected model, i.e., $\widehat{\mathbb{E}}(|\widehat{\mathcal{S}}|/|\mathcal{S}|)$. Performance is measured by the estimation error $\widehat{MSE}= \widehat{\mathbb{E}}(\|\widehat{\bnu}-\bnu\|^2)$, and by the ranking error
\[
\widehat{RE} =
\widehat{\mathbb{E}}(\sum_{i<j}\mathbb{I}\{\operatorname{sign}(\widehat{\nu}_{ij}) \neq \operatorname{sign}(\nu_{ij})\}).
\]
The ranking error $\widehat{RE}$ is the expected number of pairwise
preference directions that are incorrectly recovered.

Three experimental scenarios are considered, see Table \ref{simulation:scenarios}. In Scenario I, the minimal model is unique, moreover the triads in the minimal model have three ticks whereas all other triads have either $0$ or $1$ ticks in $\mathcal{C}_{\bnu_{\rm cyclic}}$. In Scenario II the minimal model is again unique, all triads in the minimal model have three ticks as does the triad $(4,5,6)$ which is not present in the minimal model. Finally, in Scenario III the minimal model is not unique as $\pmb{c}_{(1,2,3)}-\,\pmb{c}_{(1,2,4)}= -\,\pmb{c}_{(1,3,4)} +\pmb{c}_{(2,3,4)}$.

\FloatBarrier
\begin{table}[!ht]
    \centering
    \caption{Simulation scenarios}
    \label{simulation:scenarios}
    \begin{tabular}{cclc}
    \hline
    Scenario & $K$ & $\bnu_{\rm cyclic}$ &  \\ \hline
    I & 6, 10, 20, 50 & $ \pmb{c}_{(1,2,3)}- \pmb{c}_{(1,4,5)}$ & \\
    II & 6, 10, 20, 50 & $ \pmb{c}_{(1,4,5)}- \pmb{c}_{(2,5,6)}+\pmb{c}_{(3,4,6)}$ & \\
    III & 6, 10, 20, 50 & $\pmb{c}_{(1,2,3)}-\pmb{c}_{(1,2,4)}+ \pmb{c}_{(1,2,5)}$ &  \\  \hline
    \end{tabular}
\end{table}
\FloatBarrier

We consider only complete comparison graphs in which $n_{ij}=m$ for $K\in\{6, 10, 20, 50\}$. For Scenarios I and II we report on $\widehat{\mathbb{P}}(\mathcal{S}\subset \widehat{\mathcal{S}})$,  the empirical probability that the correct triads were selected, and on $\widehat{\mathbb{E}}(|\widehat{\mathcal{S}}|/|\mathcal{S}|)$ the standardized mean size of the selected model; the closer $\widehat{\mathbb{E}}(|\widehat{\mathcal{S}}|/|\mathcal{S}|)$ is to unity the better. For Scenario III we report on $\widehat{\mathbb{P}}(\bnu_{\rm cyclic}\in {\rm span}(\widehat{\mathcal{S}}))$, the empirical probability that the space spanned by selected triads contains true $\bnu_{\rm cyclic}$, and on $\widehat{\mathbb{E}}(|\widehat{\mathcal{S}}|/|\mathcal{S}|)$ described above. All results are based on $10^3$ simulation runs and displayed in Figures \ref{fig:1&2} and \ref{fig:3}. LASSO is implemented by cross-validation using the R package \texttt{glmnet}. All tests are carried out at the \(0.05\) level.

\begin{figure}[!htb]
    \centering
    \input{Plot_Scenarios_1_and_2.tex}
    \caption{Performance metrics for Scenarios I and II.}
    \label{fig:1&2}
\end{figure}

\begin{figure}[!htb]
    \centering
    \input{Plot_Scenario_3}
    \caption{Performance metrics for Scenarios III.}
    \label{fig:3}
\end{figure}

For scenarios I and II, when $m$ is large $\widehat{\mathcal{S}}\supset \mathcal{S}$ with high probability. The coverage probability of the LASSO, i.e., $\widehat{\mathbb{P}}(\mathcal{S}\subset \widehat{\mathcal{S}})$, is generally higher than FTBS; however, it performs poorly in terms of the average size of the selected model. For example, under Scenario I with $K=10$ and $m=30$, the LASSO outputs models of average size $\approx 3$ when the true minimal model is of size $1$. This is not surprising, since the LASSO is designed to minimize a penalized least--squares criterion with an \(L_1\) penalty, rather than to optimize support recovery directly. The FTBS output model is generally close to $1$, but the probability of correct selection is poor for small $m$. The LASSO+FTBS procedure gives a balance between LASSO and FTBS. For Scenario III, where the minimal model is not unique, the number of significant cyclic triads increases with $K$. These triads are also highly correlated because of the underlying topology. In this setting, FTBS exploits the block structure of the problem and therefore outperforms LASSO both in terms of selection probability and in terms of the average size of the selected model. This is consistent with the known selection inconsistency of the LASSO when covariates are highly correlated; see, e.g., Zhao and Yu (2006).

Finally, we examine the $\widehat{MSE}$ and $\widehat{RE}$ incurred when estimating the preference profile $\bnu$ in the three scenarios presented in Table \ref{simulation:scenarios}. We do so in six settings: $(i)$ when the true model is known in advance; $(ii)$ when the true model is unknown but selected using LASSO; $(iii)$ when the true model is unknown but selected using FTBS; $(iv)$ when the true model is unknown but selected using LASSO+FTBS; $(v)$ when the true model is unknown and the full model is fit; and $(vi)$ when the true model is unknown and the reduced (transitive) model is fit. Computing $\widehat{RE}$, requires the specification of $\bnu_{\rm linear}$, which we define as $\nu_{{\rm linear}, ij}=(i-j)/10$.

\begin{table}[htb!]
\centering
\caption{Performance measures under Scenarios I--III with $K=20$ and $m=30$. }
\label{sim:mse:table}
\resizebox{\textwidth}{!}{ % Uncomment this and the closing brace below if the table is too wide for your page margins
\begin{tabular}{l ccc ccc ccc}
\toprule
& \multicolumn{3}{c}{Scenario I} & \multicolumn{3}{c}{Scenario II} & \multicolumn{3}{c}{Scenario III} \\
\cmidrule(lr){2-4} \cmidrule(lr){5-7} \cmidrule(lr){8-10}
Method & $\widehat{MSE}$ & $\widehat{RE}$ & $\widehat{\mathbb{E}} (|\widehat{\mathcal{S}}|/|\mathcal{S}|)$ & $\widehat{MSE}$ & $\widehat{RE}$ & $\widehat{\mathbb{E}} (|\widehat{\mathcal{S}}|/|\mathcal{S}|)$ & $\widehat{MSE}$ & $\widehat{RE}$ & $\widehat{\mathbb{E}} (|\widehat{\mathcal{S}}|/|\mathcal{S}|)$ \\
\midrule
True       & 0.6935 & 0.7200 & - & 0.7207 & 0.6810 & - & 0.7479 & 0.7330 & - \\
LASSO      & 1.2901 & 0.8120 & 4.20 & 1.5408 & 0.8740 & 4.14 & 1.8485 & 1.3640 & 6.65 \\
FTBS       & 0.7198 & 0.7390 & 1.00 & 0.7598 & 0.7030 & 1.01 & 0.7774 & 0.7460 & 1.00 \\
LASSO+FTBS & 1.8688 & 1.9450 & 2.97 & 2.1856 & 2.2150 & 2.76 & 2.4845 & 2.7190 & 3.22 \\
Full       & 6.3490 & 8.0830 & --   & 6.3165 & 8.0230 & --   & 6.3501 & 8.1220 & --   \\
Reduced    & 6.6303 & 3.7580 & - & 9.6222 & 4.7300 & - & 7.6463 & 3.8130 & - \\
\bottomrule
\end{tabular}
}
\end{table}

As expected, both $\widehat{MSE}$ and $\widehat{RE}$ are minimized when the
true model specified in Table~\ref{simulation:scenarios} is known. When the
true model is unknown and selected using FTBS, the increase in
$\widehat{MSE}$ is modest, on the order of $7\%$. FTBS also outperforms the
LASSO and LASSO+FTBS in terms of both $\widehat{MSE}$ and $\widehat{RE}$.
This is consistent with the fact that FTBS is based on a sum--of--squares
criterion, which is closely aligned with the estimation error being evaluated. In comparison, fitting the full model
$(\widehat{\bnu}=\bar{\bS}_{\rm ALL})$ leads to a substantial increase in
$\widehat{MSE}$, approximately $1000\%$. A comparable increase, again nearly
$1000\%$, is observed when the reduced linearly transitive model is fitted.
This is expected, since under misspecification by a linearly transitive model the full preference profile $\bnu$ cannot be consistently recovered. The ranking error $\widehat{RE}$ exhibits the same qualitative pattern. These results show that the proposed methodology substantially improves estimation of the underlying preference profile and, consequently, can improve the induced pairwise decisions. An application to betting and wagering is presented in the next section.

\begin{summary}
The simulation results are broadly consistent with the asymptotic theory.
However, they reveal more nuanced behavior in finite samples. Specifically,
although detecting intransitivity is possible even for small $m$, see Singh
and Davidov (2026), identifying the exact cyclic structure, i.e., support
recovery, remains challenging unless $m$ is large. The LASSO tends to have
high coverage but often does so by selecting substantially larger models. In
contrast, FTBS is more parsimonious, although its exact selection probability may be low when $m$ is small. The hybrid LASSO+FTBS procedure provides a useful compromise for moderate $m$, combining the screening ability of LASSO with the parsimony of FTBS. For large $m$, standalone pruned FTBS performs well both in terms of recovery and model size. The performance measures show a similar pattern: FTBS yields small estimation error and ranking error, and substantially improves on both the full model and the reduced linearly transitive model. Thus, recovering a parsimonious cyclic structure is useful not only for structural interpretation, but also for estimating the preference profile and recovering the induced pairwise decisions.
\end{summary}

%%%%%%%%%%%%%%%%%%%%%%%%%%%%
%%%%%%%%%%%%%%%%%%%%%%%%%%%%
%%%%%%%%%%%%%%%%%%%%%%%%%%%%

\section{Illustrative example} \label{section:real:example}

Sporting competitions are an important source of interest, study and entertainment for the general public, as well as an active arena in which statistical analyses and ranking methods are passionately applied. To date, methods employed for ranking sports teams have assumed linear transitive models (Sinuany-Stern 1988, Cassady et al. 2005, Langville and Meyer 2012, Barrow et al. 2013). Alternatives allowing for cyclicalities and intransitivities were simply not available.  

We apply the framework developed in this communication to analyze data from the English Premier League (EPL), one of the foremost soccer/football leagues in the world. We focus on the 2022--23 season in which $20$ teams competed. The EPL operates on a double round--robin format, i.e., each team plays every other team twice, once on each home field. Thus, $38$ matches are played by each team resulting in a total of $380$ matches. 
In the following, the outcome of each game is defined as the difference in expected goals (xG) scored; see Roccetti et al. (2024) for a discussion of xG in evaluating soccer performance. 
The second and third columns of Table \ref{real:example:table} report the names of the teams and their positions, that is, their official ranking as it appears on the EPL website \texttt{www.premierleague.com}. Teams are alphabetically listed, and the first column of Table \ref{real:example:table} assigns each team an index number by which we shall refer to them.   

\begin{table}[!ht]
\centering
\caption{Names, positions, merits ($\widehat{\mu}_{s,i}$), Merit--Rank ($r(\widehat{\mu}_{s,i})$), Dominance--Score ($\mu_i^{**}$) and Dominance--Rank($r(\mu_i^{**})$) of teams  in English Premier League (EPL) season 2022-23} \label{real:example:table}
\begin{tabular}{rlccccc} \hline
{Index} & Name & Position & {$\widehat{\mu}_{s,i}$} & $r(\widehat{\mu}_{s,i})$ & $\mu_i^{**}$ & $r(\mu_i^{**})$\\ \hline
1 & Arsenal & 2 & 0.7325 & 3 & 17 & 2 \\
2 & Aston Villa & 7 & -0.0575 & 9 & 10 & 11 \\
3 & Bournemouth & 15 & -0.635 & 20 & 4 & 19 \\
4 & Brentford & 9 & 0.165 & 8 & 13 & 6 \\
5 & Brighton & 6 & 0.585 & 4 & 15 & 5 \\
6 & Chelsea & 12 & -0.07 & 10 & 9 & 12 \\
7 & Crystal Palace & 11 & -0.2225 & 12 & 6 & 14 \\
8 & Everton & 17 & -0.505 & 16 & 5 & 15 \\
9 & Fulham & 10 & -0.4325 & 14 & 7 & 13 \\
10 & Leeds United & 19 & -0.49 & 15 & 4 & 17 \\
11 & Leicester City & 18 & -0.3225 & 13 & 10 & 8.5 \\
12 & Liverpool & 5 & 0.54 & 5 & 16 & 3 \\
13 & Manchester City & 1 & 1.16 & 1 & 18 & 1 \\
14 & Manchester Utd & 3 & 0.43 & 6 & 12 & 7 \\
15 & Newcastle Utd & 4 & 0.805 & 2 & 15 & 4 \\
16 & Nott'ham Forest & 16 & -0.6175 & 19 & 4 & 18 \\
17 & Southampton & 20 & -0.5825 & 17 & 5 & 16 \\
18 & Tottenham & 8 & 0.195 & 7 & 10 & 8.5 \\
19 & West Ham & 14 & -0.095 & 11 & 10 & 10 \\
20 & Wolves & 13 & -0.5825 & 18 & 3 & 20 \\
\hline
\end{tabular}
\end{table}

First, ignoring possible cyclicalities, the data is analyzed assuming the reduced model, i.e., assuming the relation \eqref{nu.ij=mu.i-mu.j} holds. The fourth column of Table \ref{real:example:table} reports the estimated merits and the fifth column the merit--based rankings where the rank of the $i^{th}$ team is given by $r(\widehat{\mu}_i)=\sum_{j\neq i} \mathbb{I}(\widehat{\mu}_i>\widehat{\mu}_j)$. For example, Manchester City is ranked first with a merit of $1.16$. In addition, note that merit--based ranking coincides with both the row--sum ranking method and the method proposed by Saari (2014) and Saari (2021), see Equation \eqref{eq.saari.score} in the Supplement.  

We find that the $p$--value for lack of fit of the reduced model is
$<10^{-3}$. However, $m=2$, so identifying the minimal model is challenging.
The LASSO shrinks all cyclic parameters to zero and returns the baseline
transitive model, thereby also precluding the use of the LASSO+FTBS approach. For FTBS, screening by the FDR procedure at level $0.1$ yields a tick--table, $\bC_{{\bnu}_{\rm cyclic}^*}$, with no $2$- or $3$-tick columns and exactly $54$ singly ticked triads. These triads are then pruned at level $\alpha=0.01$, yielding a parsimonious final model containing only three cyclic triads: \( \{(2,5,19), (6,12,14), (10,11,13)\}\). The selected model provides a parsimonious adequate fit. It has lower AIC and BIC values than its submodels, and unlike those submodels, it is not rejected by the lack--of--fit test. Thus it is the smallest model passing the lack--of--fit test; see Table~\ref{EPL-ftbs}. Furthermore, the triads identified by pruned FTBS are edge--disjoint, which supports the stability of the selected cyclic structure. The selected model is given by
$${\widehat{\bnu}}_s=\bB{\widehat{\bmu}}_s+\pmb{c}_{(2,5,19)}{\widehat{{\gamma}}}_{(2,5,19)}+
\pmb{c}_{(6,12,14)}{\widehat{{\gamma}}}_{(6,12,14)}+
\pmb{c}_{(10,11,13)}{\widehat{{\gamma}}}_{(10,11,13)}$$ 
where $\widehat{{\bmu}}_s$ and $\widehat{\gamma}$'s are estimated as described in Theorem \ref{thm:mu:lambda:estimation:properties}. The merits associated with the selected model are displayed in column four of Table \ref{real:example:table} and $\widehat{{\gamma}}_{(2,5,19)}=1.18$, ${\widehat{{\gamma}}}_{(6,12,14)}=1.1$ and ${\widehat{{\gamma}}}_{(10,11,13)}=1.07$. The resulting preference profile, \(\widehat{\bnu}_s\), is non--transitive and therefore cannot be represented by a usual ranking. This is illustrated in Figure~\ref{example:preference:graph}, where an arrow from \(i\) to \(j\) indicates that \(i \succ j\).

\begin{table}[htb!]
\centering
\caption{Model selection metrics comparing the baseline transitive model, selected by both LASSO and LASSO+FTBS, and the model selected by FTBS.}
\label{EPL-ftbs}
\begin{tabular}{lccc}
\toprule
Method & $|\widehat{\mathcal{S}}|$ & AIC & BIC \\
\midrule
Transitive model & 0 & 1118.56 & 1197.37 \\
FTBS      & 3 & 1101.39 & 1192.02 \\
\bottomrule
\end{tabular}
\end{table}

\begin{figure}[!htbp]
    \centering
\begin{tikzpicture}[
            > = stealth, % arrow head style
            shorten > = 1pt, % don't touch arrow head to node
            auto,
            node distance = 2.25cm, % distance between nodes
            semithick % line style
        ]

        \tikzstyle{every state}=[draw = black, thick,
            fill = white, minimum size = 4mm]
        \node[state] (2) {$2$};
        \node[state] (19) [above  of=2] {$19$};
        \node[state] (5) [right of=19] {$5$};
        \node[state] (18) [below of=5] {$18$};

        \path[->] (2) edge node {} (5);
        \path[->] (19) edge node {} (2);
        \path[->] (5) edge node {} (19);
        \path[->] (18) edge node {} (2);
        \path[->] (5) edge node {} (18);
        \path[->] (18) edge node {} (19);
    \end{tikzpicture}
\caption{Preferences among teams with indices $\{2,5,18,19\}$}
    \label{example:preference:graph}
\end{figure}
\FloatBarrier

Nevertheless, some applications require a ranking even in the presence of cyclicality and intransitivity, see Saari (2014) and Saari (2021). We propose a ranking based on dominance scores, defined for $i^{th}$ item as:
\begin{equation*} 
\mu_i^{**} = \sum_{j=1}^{K}  \mathbb{I}(\nu_{ij}>0).
\end{equation*}
Notice that the dominance score is reminiscent of the well known Borda--Count advocated among some voting theorists (e.g., Saari 2023). Using the dominance score rank $i$ above $j$ if $\mu_i^{**} > \mu_j^{**}$. Ties among the dominance score can be further broken by repeatedly computing \eqref{eq.dom.score} on sets of the form $\{j: \mu_i^{**}=\mu_j^{**}\}$. Columns six and seven in Table \ref{real:example:table} display the dominance scores and ranks for all teams in the EPL, respectively. We note that rankings in the presence of intransitivity are necessarily imperfect, since any global ranking method must collapse cycles into a linear order and therefore cannot fully respect the observed cyclicalities. One way of addressing and partially resolving these issues is to define a ranking with respect to each item $i$. Define the sets: (i) $S_i=\{j: \widehat{\nu}_{ij}<0\}$; (ii) $I_i=\{j: \widehat{\nu}_{ij}>0\}$; and (iii) $E_i=\{j: \widehat{\nu}_{ij}=0\}$. The elements in $S_i$ and $I_i$ are ranked above (superior) and below (inferior) item $i$, respectively, and the elements in $E_i$ are equivalent to $i$. See Table \ref{table:rank:sets} in the Supplement for the corresponding sets.
\begin{remark}
A mathematical discussion of ranking under intransitive preference models is provided in Section~\ref{section:ranking:intransitivity} of the Supplement.
\end{remark}

Finally, we explore implications for wagers and bets. It is well known that if one believes that $\mathbb{P}(W)=\omega$ for some event $W$, then a fair bet on $W$ pays $(1-\omega)/\omega$ monetary units for each monetary unit staked; see, e.g., Epstein (2012). Suppose now that a counterparty prices the event according to $\mathbb{P}(W)=\omega$, whereas the true probability is $\mathbb{P}(W)=\tau$. If $\tau>\omega$, then betting on $W$ has positive expected gain. Indeed, by staking one monetary unit, the expected gain is $\tau(1-\omega)/\omega-(1-\tau)=(\tau-\omega)/\omega$. Similarly, if $\tau<\omega$, then one should instead bet against $W$, equivalently on $W^c$. In that case, the expected gain from staking one monetary unit is $(1-\tau)\omega/(1-\omega)-\tau=(\omega-\tau)/(1-\omega)$. Thus, whenever the true probability $\tau$ differs from the probability $\omega$ used to price the bet, a positive expected gain is available by betting on the side favored by the discrepancy.

Next, let $W_{ij}$ denote the event that team $i$ wins against team $j$ in their next encounter. Assuming that $Y_{ij}$ is a $\mathcal{N}(\nu_{ij},\sigma^2)$ RV we find that
$$ \mathbb{P}(W_{ij}) =
\Phi\left(\frac{\nu_{ij}}{\sigma}\right),$$
where $\Phi$ is the standard normal cumulative distribution function. Let $\omega_{ij}$ denote the probability of $W_{ij}$ assuming LST, i.e., model \eqref{nu.ij=mu.i-mu.j}, and let $\tau_{ij}$ denote the probability of $W_{ij}$ under a possibly intransitive model, see \eqref{bnu:decomposed:linear:cyclic}. Thus,
$\omega_{ij} = \Phi(\nu_{ij,\mathrm{LST}}/{\sigma}),
$ where $\nu_{ij,\mathrm{LST}}$ is the $(i,j)^{th}$ entry of $\bnu_{\mathrm{LST}}$, as given in Theorem~\ref{cyclicality:lemma:nu:mu} (see the Supplement). Most gambling systems (e.g., Langville and Meyer, 2012) incorrectly assume that $\bnu_{\mathrm{LST}}$ represents the true value of $\bnu$. In reality, the true probability is $\tau_{ij} = \Phi(\nu_{ij,\mathrm{TRU}}/{\sigma}),
$ where $\bnu_{\mathrm{TRU}}$ is the true underlying parameter vector.

The expected gain from betting one monetary unit for, or against, the event $W_{ij}$ is then given by
\[
\mathrm{Win}_{ij} =
\begin{cases}
\displaystyle \frac{\tau_{ij} - \omega_{ij}}{\omega_{ij}}, & \text{if } \tau_{ij} > \omega_{ij}, \\[2ex]
\displaystyle \frac{\omega_{ij} - \tau_{ij}}{1 - \omega_{ij}}, & \text{if } \tau_{ij} < \omega_{ij}, \\[2ex]
0, & \text{if } \tau_{ij} = \omega_{ij}.
\end{cases}
\]
Accordingly, the total expected gain over a round--robin tournament (where each pair $(i,j)$ meets once) is
$$ \mathrm{TotalWin} = \sum_{1 \leq i < j \leq K}
\left[
  \frac{\tau_{ij} - \omega_{ij}}{\omega_{ij}}\,\mathbb{I}(\tau_{ij} > \omega_{ij})
  + \frac{\omega_{ij} - \tau_{ij}}{1 - \omega_{ij}}\,\mathbb{I}(\tau_{ij} < \omega_{ij})
\right]. $$
This expression quantifies the cost of belief miscalibration due to model misspecification, i.e.,  assuming transitivity when cyclicality may be present. As an illustration, we estimate $\mathrm{Win}_{ij}$ and $\mathrm{TotalWin}$ using EPL2023 data. Specifically, we set $\nu_{ij,\mathrm{LST}} = \widehat{\mu}_i - \widehat{\mu}_j$, $\nu_{ij,\mathrm{TRU}} = \widehat{\nu}_{ij,s}$, and $\sigma = \widehat{\sigma}$ as given in Proposition~\ref{prop:H0:cyclic:triad} (see the Supplement). Notably, $\mathrm{Win}_{ij} > 0$ only if $(i, j) \in \operatorname{support}(\mathcal{C}_{\gamma_s})$; hence in the example at hand, $\mathrm{Win}_{ij} > 0$ whenever $i, j \in \{2, 5, 6, 10, 11, 12, 13, 14, 19\}$, and zero otherwise. In particular, $\mathrm{Win}_{2,5} = 2.67$, $\mathrm{Win}_{2,19} = 0.93$, $\mathrm{Win}_{5,19} = 0.23$, $\mathrm{Win}_{6,12} =2.37$, $\mathrm{Win}_{6,14} =0.32$, $\mathrm{Win}_{10,11} =1.17$, $\mathrm{Win}_{10,13} =0.02$, $\mathrm{Win}_{11,13} =11.08$, and $\mathrm{Win}_{12,14} =0.69$ so that 
$$\mathrm{TotalWin} = 19.5.$$
In summary, modeling cyclicality in paired comparison data can have a considerable impact on betting and wagering systems. 

%%%%%%%%%%%%%%%%%%%%%%%%%%%%
%%%%%%%%%%%%%%%%%%%%%%%%%%%%
%%%%%%%%%%%%%%%%%%%%%%%%%%%%

\section{Discussion} \label{discussion}

Traditional approaches to paired comparison data often rely on transitivity, representing preferences through a global merit vector. This paper develops a framework for modeling paired comparison data in the presence of cyclicality and intransitivity. The key idea is the orthogonal decomposition of the parameter space $\mathcal{N}$ into a linear transitive subspace and a cyclic subspace, leading to the representation \eqref{bnu:decomposed:linear:cyclic}. This representation separates merit--based effects from cyclicality and provides a basis for model selection, estimation and interpretation. The decomposition is related to, but distinct from, earlier representations of paired comparison data. Scheffé's ANOVA decomposition, \(\nu_{ij}=\mu_i-\mu_j+\gamma_{ij}\), with suitable row and column constraints on the interaction terms, yields a saturated model and therefore permits consistent estimation of any $\bnu\in\mathcal{N}$. However, it does not exploit the intrinsic algebraic structure of paired comparison data: the merit and interaction terms do not correspond directly to linear transitivity and cyclicality. Saari (2014, 2021) obtained bases for the linear and cyclic subspaces and
used them to study ranking. His work was structural rather than inferential.
A related perspective was developed by Jiang et al. (2011), who used the
language of combinatorial Hodge theory. Recent statistical work has built on
this viewpoint: Okahara et al. (2026) select a basis from the cyclic dictionary and apply Bayesian methods, while Lee and Chen (2026) study sparse representations of the preference profile when viewed as an antisymmetric matrix. The present paper differs in that it models the preference profile directly and develops procedures for recovering a minimal cyclic representation.

The recovery of a minimal cyclic representation results in a model selection problem. Although $L_1$--regularized methods such as the LASSO are natural candidates, exact support recovery requires conditions, such as mutual incoherence, that need not hold for the highly dependent cyclic dictionary. This motivates the proposed FTBS procedure. The simulations show that LASSO tends to provide high coverage at the cost of selecting larger models, whereas FTBS and its pruned version favor more parsimonious representations. The empirical example further illustrates that explicitly modeling cyclic structure can improve estimation of the preference profile and affect ranking, prediction, and wagering decisions.

Several extensions are natural. First, the computational and statistical performance of FTBS can be improved by refining the block structure, combining forward tick--based selection with backward deletion, or incorporating information criteria such as AIC and BIC (Claeskens and Hjort, 2008). Second, while this paper focuses on complete PCGs, many applications involve incomplete comparison graphs. Extending the decomposition, estimation, and recovery theory to incomplete PCGs is an important direction. Third, longer cycles of the form \(i_1\succeq i_2\succeq\cdots\succeq i_r\succeq i_1\) can be represented within the cyclic subspace, since such cycles can be expressed as linear combinations of cyclic triads. Developing model selection procedures that directly account for longer cycles may lead to alternative interpretable representations of cyclic structure. Another important extension concerns binary, ordinal, and other non--cardinal paired comparison data. Although this paper focuses on cardinal PCD, the proposed framework can be extended to such settings. For example, in binary PCD the probability that item $i$ is preferred over item $j$ is typically modeled as $\mathbb{P}(Y_{ijk}=1)=F(\mu_i-\mu_j)$ where $F$ is a distribution function, symmetric about $0$, i.e., $F(x)+F(-x)=1$ for all $x\in\mathbb{R}$, and $\mu_1,\ldots,\mu_K$ are the merit parameters, cf. Oliveira et al. (2018). Specifically, when $F$ is the standard logistic distribution we obtain the relation
\begin{equation*} \label{Eq.BT.model}
\mathbb{P}(Y_{ijk}=1) = \frac{\exp(\mu_i-\mu_j)}{1+\exp(\mu_i-\mu_j)}.    
\end{equation*}
This is the well--known Bradley--Terry model (BT) (Bradley and Terry 1952, Hunter 2004, Cattelan 2012) whereas when $F$ is the standard normal DF we obtain the so--called Thurstone model (Thurstone 1927, Böckenholt, 2006). Clearly, the classic BT model can be extended. Modelling cyclicalities is straightforward; instead of $\mu_i-\mu_j$ write $\nu_{ij}$ and conduct inference on $\bnu\in \mathcal{N}$. A bit of reflection shows that the modeling framework for cardinal PCD, discussed in Section \ref{section:nu:decomposition}, is applicable. The estimators and corresponding tests will, of course, be different; a detailed analysis is clearly warranted. 

In summary, cyclicality is not merely a lack of fit of the transitive ranking model. It is a structural component of the preference profile that can be explicitly modeled, selected, estimated, and interpreted. Moreover, recovering the minimal cyclic model yields a parsimonious description of the preference profile and can improve ranking, prediction, and decision making in applied settings. The proposed framework, therefore, changes the role of cyclicality in paired comparison analysis: from an unexplained anomaly to an identifiable and decision--relevant feature of the data.

%%%%%%%%%%%%%%%%%%%%%%%%%%%%
%%%%%%%%%%%%%%%%%%%%%%%%%%%%
%%%%%%%%%%%%%%%%%%%%%%%%%%%%

\section*{Acknowledgments}

The work of Rahul Singh was conducted while a post--doctoral fellow at the University of Haifa. The work of Ori Davidov was partially supported by the Israeli Science Foundation Grant No. 2200/22 and gratefully acknowledged. 

{\setstretch{1}

}

\clearpage

\appendix
\begin{center}
{\bf\Large Supplement to \\ ``Modeling cyclicality and intransitivity in paired comparisons data''}
\end{center}
\pagenumbering{arabic} % Ensures it's 1, 2, 3...
\setcounter{page}{1}
\counterwithin{equation}{section} 

The supplement is divided into five sections. Section A contains the proofs of all theorems and propositions presented in the main text. Section B provides the relationship between the reduced and full models and Section C provides some discussion and results on ranking methodologies under intransitivity. Section D provides a goodness of fit test. Section E presents some numerical results related to the illustrative example.

\section{Proofs} \label{app:proofs}

Section A contains the proofs of all theorems and propositions appearing in the main text. Some supporting auxiliary remarks and lemmas are also provided.  

\subsubsection*{Proof of Theorem \ref{thm:unique:minimal:model}:}
\begin{proof}
 By Definition if \ref{def:minimal:model} $\bnu_{\rm cyclic}\in {\rm span}(\mathcal{C}_{\pmb{\gamma}_1})$ and $\bnu_{\rm cyclic}\in {\rm span}(\mathcal{C}_{\pmb{\gamma}_2})$ and consequently $\bnu_{\rm cyclic}\in {\rm span}(\mathcal{C}_{\pmb{\gamma}_1}) \cap {\rm span}(\mathcal{C}_{\pmb{\gamma}_2})$. Next suppose that $ {\rm span}(\mathcal{C}_{\pmb{\gamma}_1}) = {\rm span}(\mathcal{C}_{\pmb{\gamma}_2})$. It immediately follows that $\mathcal{C}_{\pmb{\gamma}_1}$ and $\mathcal{C}_{\pmb{\gamma}_2}$ have same set of items, say $\{1,\ldots,L\}$. For if $i$ is present in $\mathcal{C}_{\pmb{\gamma}_1}$ and absent in $\mathcal{C}_{\pmb{\gamma}_2}$ then there exist a triad $\pmb{c}_{(i,j,k)}\in {\rm span}(\mathcal{C}_{\pmb{\gamma}_1})$ and $\pmb{c}_{(i,j,k)}\notin {\rm span}(\mathcal{C}_{\pmb{\gamma}_2})$ for all $(j,k)\notin {\rm support}(\mathcal{C}_{\pmb{\gamma}_2})$. Hence $ {\rm span}(\mathcal{C}_{\pmb{\gamma}_1}) \neq {\rm span}(\mathcal{C}_{\pmb{\gamma}_2})$. Next, suppose that $(i,j)\in {\rm support}(\mathcal{C}_{\pmb{\gamma}_1})$. Then there exists a $k\leq L$ such that $\pmb{c}_{(i,j,k)}\in {\rm span}( \mathcal{C}_{\pmb{\gamma}_1})$ and consequently $\pmb{c}_{(i,j,k)}\in {\rm span}( \mathcal{C}_{\pmb{\gamma}_2})$. Therefore if the spans are equal so are the supports.

Let $\pmb{\gamma}_1$ and $\pmb{\gamma}_2$ be two minimal models. By assumption $(i,j)\in{\rm support}(\mathcal{C}_{\pmb{\gamma}_1})$ implies $\nu_{{\rm cyclic},ij}\neq 0$, therefore it also follows that $(i,j)\in{\rm support}(\mathcal{C}_{\pmb{\gamma}_2})$ and consequently ${\rm support}(\mathcal{C}_{\pmb{\gamma}_1})\subseteq{\rm support}(\mathcal{C}_{\pmb{\gamma}_2})$. Now if ${\rm support}(\mathcal{C}_{\pmb{\gamma}_1})\subset{\rm support}(\mathcal{C}_{\pmb{\gamma}_2})$, there exists some $(i,j)\in {\rm support}(\mathcal{C}_{\pmb{\gamma}_2})$ and $(i,j)\notin {\rm support}(\mathcal{C}_{\pmb{\gamma}_1})$. Thus there is some triad $(i,j,t)$ which belongs to $\mathcal{C}_{\pmb{\gamma}_2}$ but not $\mathcal{C}_{\pmb{\gamma}_1}$ consequently $s(\pmb{\gamma}_2)>s(\pmb{\gamma}_1)$ and therefore $\pmb{\gamma}_2$ can not be a minimal model. Thus ${\rm support}(\mathcal{C}_{\pmb{\gamma}_1})={\rm support}(\mathcal{C}_{\pmb{\gamma}_2})$ and completing the proof of third part. 

Finally suppose that any two triads in a minimal model $\pmb{\gamma}$ spanned by $\mathcal{C}_{\pmb{\gamma}}$ share at most one index, i.e., if $\pmb{c}_{(i,j,k)},\pmb{c}_{(u,v,w)} \in \mathcal{C}_{\pmb{\gamma}}$ then $|\{i,j,k\}\cap \{u,v,w\}|\le 1$. Using \eqref{cyclic:triangle:not:unique}  it is easily verified that such triads are associated with orthogonal columns of $\bC$. It is also immediate that if $(i,j)\in {\rm support}(\mathcal{C}_{\pmb{\gamma}})$ then $(i,j)$ belongs to a single triad in $\mathcal{C}_{\pmb{\gamma}}$. By definition
\begin{align} \label{eq:nu_cyclic_components}
\bnu_{{\rm cyclic}}= \sum_{\pmb{c}_{(i,j,k)}\in\mathcal{C}_{\pmb{\gamma}}} \pmb{c}_{(i,j,k)}\pmb{\gamma}_{ijk}.   
\end{align}
It is clear that if $(s,t)\notin {\rm support}(\mathcal{C}_{\pmb{\gamma}})$ then $\nu_{{\rm cyclic},st}=0$, whereas if $(s,t)\in {\rm support}(\mathcal{C}_{\pmb{\gamma}})$ then $(s,t)$ belongs to a single triad $(i,j,k)$, say, and $\nu_{{\rm cyclic},st}={c}_{(i,j,k),st}{\gamma}_{ijk} \neq 0$. We conclude that $\nu_{{\rm cyclic},ij}\neq0$  if and only if $(i,j)\in{\rm support}(\mathcal{C}_{\pmb{\gamma}})$. Also $\sum_{1\leq i<j\leq K} \mathbb{I}(\nu_{{\rm cyclic},ij}\neq0) = 3 s$, where $s$ is size of the model. Next suppose there exist a minimal model $\pmb{\gamma}'$ for which $\pmb{\gamma}\neq \pmb{\gamma}'$. 

It follows from the discussion above that if $\bnu_{{\rm cyclic},ij}\neq 0$ then $(i,j) \in {\rm support}(\mathcal{C}_{\pmb{\gamma}'})$. These pairs, however, are in ${\rm support}(\mathcal{C}_{\pmb{\gamma}})$ and therefore ${\rm support}(\mathcal{C}_{\pmb{\gamma}})\subseteq {\rm support}(\mathcal{C}_{\pmb{\gamma}'})$. Remark \ref{triads:from:support} implies that $s({\pmb{\gamma}})< s({\pmb{\gamma}'})$, i.e., the size of the model $\pmb{\gamma}'$ is larger than the size of the model $\pmb{\gamma}$. This leads to a contradiction and the conclusion that $\pmb{\gamma}=\pmb{\gamma}'$.
\end{proof}

\begin{remark} \label{triads:from:support}
If a minimal model has to be constructed using a given support, then we seek a minimum number of triads using the support, such that these triads are linearly independent. That is, if $(i,j),(i,k),(j,k)$ are present in the support then $\pmb{c}_{(i,j,k)}$ is a viable triad and we find such independent triads until support is covered. Now if triads in a minimal model $\pmb{\gamma}$ share at most one index then any comparison appears in only one triad of $\mathcal{C}_{\pmb{\gamma}}$. Precisely there is a one-to-one mapping between comparisons and triads in $\pmb{\gamma}$. Consequently only triads in $\mathcal{C}_{\pmb{\gamma}}$ can be made using ${\rm support}(\mathcal{C}_{\pmb{\gamma}})$, therefore the (minimum) number of triads made using ${\rm support}(\mathcal{C}_{\pmb{\gamma}})$ is $s({\pmb{\gamma}})$.
\end{remark}

\subsubsection*{Proof of Theorem \ref{thm:ticks:real:model}:}
\begin{proof}
Suppose that a minimal model $\pmb\gamma$ has a single one $0$-tick triad $(i,j,k)$. It follows that all other triads in $\mathcal{C}_{\pmb{\gamma}}$ have at least one tick and that $\bnu_{{\rm cyclic}, ij}= \bnu_{{\rm cyclic}, ik}=\bnu_{{\rm cyclic}, jk}= 0$. Now by \eqref{eq:nu_cyclic_components} for any pair $(s,t)$
\begin{align}\label{eq:0:nucyclic}
\nu_{{\rm cyclic},st}=\sum_{(u,v,w)\in \mathcal{C}_{\pmb{\gamma}}} {c}_{(u,v,w),st}{\gamma}_{uvw}.
\end{align}
Therefore if $\bnu_{{\rm cyclic},ij}=0$ then there must be some other triad in which the comparison $(i,j)$ appears. Similarly for the pairs $(j,k)$ and $(i,k)$. Also notice that any two distinct triads can have at most one comparison in common and therefore the minimal model will have at least three more triads with two indices in common with $(i,j,k)$ and having at least $1$--tick. Assume that there are exactly three such triads. These must be $\pmb{c}_{(i,j,s)},\pmb{c}_{(j,k,t)}, \pmb{c}_{(i,k,u)}$ for some $s,t$ and $u$. By \eqref{eq:0:nucyclic}, $\nu_{{\rm cyclic},ij}= {c}_{(i,j,k),ij}{\gamma}_{ijk}+ {c}_{(i,j,s),ij}{\gamma}_{ijs}=0$ implying that ${\gamma}_{ijk}=-{\gamma}_{ijs}$. A similarly relation holds for $\nu_{{\rm cyclic},jk}= 0$ and $\nu_{{\rm cyclic},ik}= 0$ and therefore ${\gamma}_{ijk}=-{\gamma}_{jkt}$ and ${\gamma}_{ijk}={\gamma}_{iku}$. Combining these equalities we find that ${\gamma}_{ijk}=-{\gamma}_{ijs}=-{\gamma}_{jkt}= {\gamma}_{iku}$. It is now clear that the elements of the vector $\pmb{c}_{(i,j,k)}-\pmb{c}_{(i,j,s)}-\pmb{c}_{(j,k,t)}+\pmb{c}_{(i,k,u)}$ which correspond to comparisons among the items $\{i,j,k\}$ are $0$. This implies by \eqref{eq:0:nucyclic} that  
$$\bnu_{\rm cyclic}= \bnu_{\rm cyclic}^{(+ijk)} + \bnu_{\rm cyclic}^{(-ijk)},$$
where $\bnu_{\rm cyclic}^{(+ijk)}= {\gamma}_{ijk} (\pmb{c}_{(i,j,k)}-\pmb{c}_{(i,j,s)}-\pmb{c}_{(j,k,t)}+\pmb{c}_{(i,k,u)})$ 
is the linear combination of the triads involving comparisons of the items in $\{i,j,k\}$ and $\bnu_{\rm cyclic}^{(-ijk)}= \bnu_{\rm cyclic}- \bnu_{\rm cyclic}^{(+ijk)}$. 

Since all other triads in the minimal model have at least one tick we assume without any loss of generality that $\bnu_{{\rm cyclic}, js}, \bnu_{{\rm cyclic}, kt}\neq 0$. Using \eqref{cijk:relation:with:other:c} we find 
\begin{align*} 
\pmb{c}_{(i,j,k)}-\pmb{c}_{(i,j,s)}= \pmb{c}_{(i,j,k)}-\pmb{c}_{(i,j,s)} + \pmb{c}_{(i,k,t)}- \pmb{c}_{(i,k,t)} = \pmb{c}_{(j,k,s)} -\pmb{c}_{(i,k,t)}.
\end{align*}
Thus by using the above relation, we can rewrite  $\bnu_{\rm cyclic}^{(+ijk)}$ as
$$\bnu_{\rm cyclic}^{(+ijk)}= {\gamma}_{ijk} ( \pmb{c}_{(j,k,s)} -\pmb{c}_{(i,k,t)} -\pmb{c}_{(j,k,t)}+\pmb{c}_{(i,k,u)}).$$
Hence we obtain another minimal model, say $\pmb{\gamma}'$, in which there are no triads with $0$ ticks. If there are more than three triads with two indices in common with $\{i,j,k\}$ then we can write  $\bnu_{\rm cyclic}^{(+ijk)}= \alpha_1(\pmb{c}_{(i,j,k)}-\pmb{c}_{(i,j,s_1)}-\pmb{c}_{(j,k,t_1)}+\pmb{c}_{(i,k,u_1)})+\ldots+\alpha_l(\pmb{c}_{(i,j,k)}-\pmb{c}_{(i,j,s_l)}-\pmb{c}_{(j,k,t_l)}+\pmb{c}_{(i,k,u_l)})$. Applying the preceding arguments to each term appearing in $\bnu_{\rm cyclic}^{(+ijk)}$ separately we arrive at the same conclusion, i.e., there exist an alternative minimal model with no $0$--tick triads. Finally, if there are $l>1$ $0$--tick triads in a minimal model then we repeat the procedure outline above for each one of them. This completes the proof of the first part of Theorem \ref{thm:ticks:real:model}.

Suppose that any two triads in a minimal model $\pmb{\gamma}$ share at most one index. The proof of Theorem \ref{thm:unique:minimal:model} shows that if $(i,j) \in {\rm support}(\mathcal{C}_{\pmb{\gamma}})$ then $\bnu_{{\rm cyclic},ij}\neq 0$. This immediately implies that all triads in $\pmb{\gamma}$ have $3$--ticks. Next it is shown, by example, that a triad $(i,j,k)$ may have $3$-ticks yet not be a member of any minimal model. Let 
$$\bnu_{\rm cyclic}=\lambda_{1} \pmb{c}_{(i,j,s)}+ \lambda_{2}\pmb{c}_{(j,k,t)}+\lambda_{3}\pmb{c}_{(i,k,u)},$$
where $|\lambda_l|$'s are distinct. It is not hard to see that the minimal model is unique. Also note that the triads $(i,j,s),~ (j,k,t)$ and $(i,k,u)$ all have $3$ ticks. However so does the triad $(i,j,k)$. This completes the proof of the second part of the theorem.    

Finally consider $\alpha\in\mathbb{R}\setminus\{0\}$ and set   
$$\bnu_{\rm cyclic}= \pmb{c}_{(1,2,3)}-\alpha\pmb{c}_{(1,2,4)}- (1-\alpha)\pmb{c}_{(1,2,5)}.$$
It is easy to verify that the minimal model for $\bnu_{\rm cyclic}$ above is unique. It is straightforward to verify that all triads in the minimal model have two ticks. This completes the proof. 
\end{proof}

\subsubsection*{Proof of Theorem \ref{thm:mu:lambda:estimation:properties}:}

\begin{proof}The proof is analogous to that of Theorem 4.2 in Singh et al. (2025). 
Notice that relation \eqref{joint:nu:model:simplified} can be rewritten as
\begin{align*} 
\bY = \pmb{H}\begin{pmatrix}
    \bmu\\ \pmb\gamma_{s}
\end{pmatrix} + \pmb{\epsilon},
\end{align*}
where $\pmb{H}=(\bL\bB,\,\bL\bC_{s})$. Therefore, the least squares solution to the above relation is
\[
\begin{pmatrix}
\widehat{\bmu}\\ \widehat{\pmb{\gamma}}_s
\end{pmatrix} = (\pmb{H}^{\top} \pmb{H})^{+} \pmb{H}^{\top} \bY
\]
Consequently, under the stated assumptions, the expectation of the LSE is
\begin{align*}
\mathbb{E}\begin{pmatrix}
\widehat{\bmu}\\ \widehat{\pmb{\gamma}}_s
\end{pmatrix} = (\pmb{H}^{\top} \pmb{H})^{+} \pmb{H}^{\top} \pmb{H} \begin{pmatrix}
{\bmu}\\ {\pmb{\gamma}}_s\end{pmatrix}.
\end{align*}
We have $(\b1^{\top},\pmb{0}^{\top})^{\top}\in \mathrm{ker}(\pmb{H}^{\top} \pmb{H})$ and $(\b1^{\top},\pmb{0}^{\top}) ({\bmu}^{\top}, {\pmb{\gamma}}_s^{\top})^{\top}=0$, so $ ({\bmu}^{\top}, {\pmb{\gamma}}_s^{\top})^{\top} \in \mathrm{im} (\pmb{H}^{\top} \pmb{H})$. Consequently $(\pmb{H}^{\top} \pmb{H})^{+} \pmb{H}^{\top} \pmb{H} ({\bmu}^{\top}, {\pmb{\gamma}}_s^{\top})^{\top} = ({\bmu}^{\top}, {\pmb{\gamma}}_s^{\top})^{\top}$. Standard calculations show that the LSEs have the form \eqref{mu:thm4.1} and \eqref{nu:thm4.1}.

Next, observe that 
     \begin{align*}
          \mathbb{V}ar\begin{pmatrix}
             \widehat{\bmu}\\ \widehat{\pmb{\gamma}}_s
         \end{pmatrix}= 
         (\pmb{H}^{\top} \pmb{H})^{+}\sigma^2 \b1.
     \end{align*}
Under the assumption that $\mathcal{S}$ is a minimal model,  the second smallest eigenvalue of $\pmb{H}^{\top} \pmb{H}$ diverges, and hence $(\pmb{H}^{\top} \pmb{H})^{+}\to\pmb{O}$ as $n\to\infty$. Consequently $\mathbb{V}ar((\widehat{\bmu}^{\top},\widehat{\pmb{\gamma}_s}^{\top})^{\top}\to\pmb{O}$.  We already have $\mathbb{E} (\widehat{\bmu}^{\top},\widehat{\pmb{\gamma}}_s^{\top})= ({\bmu}, {\pmb{\gamma}}_s)$. Therefore, we conclude that $\widehat{\bmu}\to\bmu$ and $\widehat{\pmb{\gamma}}_s\to\pmb{\gamma}_s$ in probability as $n\to\infty$.

Straightforward algebraic manipulations yield,
\begin{align*}
	\sqrt{n}(\begin{pmatrix}
		\widehat{\bmu}\\ \widehat{\pmb{\gamma}}_s
	\end{pmatrix} - \begin{pmatrix}
		{\bmu}\\ {\pmb{\gamma}}_s
	\end{pmatrix}) 
	= (\frac{1}{n}\pmb{H}^{\top} \pmb{H})^{+} \frac{1}{\sqrt{n}} \pmb{H}^{\top}\pmb{\epsilon}.
\end{align*}
Under our assumptions $ (\pmb{H}^{\top} \pmb{H}/n)^{+} \to \pmb{\Sigma}_{\mu,\gamma_s} $ as $n\to\infty$. Observe that for $\pmb{a}\not\in \mathrm{ker}(\pmb{H})$ we have
$$\pmb{a}^{\top}\pmb{H}^{\top} \pmb{\epsilon} /\sqrt{n}= \sum_{i=1}^{n}d_i\epsilon_i,$$ 
where $d_i$ and $\epsilon_i$ are the $i^{th}$ components of $\pmb{a}^{\top}\pmb{H}^{\top} /\sqrt{n}$ and $\pmb{\epsilon}$ respectively. Further note that $d_i=\pmb{a}^{\top} \pmb{h}_i$ where $\pmb{h}_i$ is the $i^{th}$ row of  $\pmb{H}$. 
Using the Rayleigh's theorem (e.g., Horn and Johnson 2007, Theorem 4.2.2, p. 234), we have
\begin{align*}
\sum_{i=1}^{n} d_i^{2} =~
 \pmb{a}^{\top} (\pmb{H}^{\top} \pmb{H}) \pmb{a} 
\geq~ \pmb{a}^{\top}\pmb{a}\, \lambda_2(\pmb{H}^{\top} \pmb{H}).
\end{align*}
Now by the Cauchy--Schwartz inequality
\begin{align*}
	d_i^{2} = (\pmb{a}^{\top} \pmb{h}_i)^{2} \leq \pmb{a}^{\top}\pmb{a}\, \max_{i} \pmb{h}_i \pmb{h}_i^{\top}.
\end{align*}
Consequently, due to the construction of $\pmb H$, we have 
\begin{align*}
	\max_{1\leq i\leq n}\frac{d_i^{2}}{\sum_{i=1}^{n} d_i^{2}} \leq 
	\frac{ \max_{i} \pmb{h}_i \pmb{h}_i^{\top} }{\lambda_2(\pmb{H}^{\top} \pmb{H})} \to 0 \text{ as } n\to\infty.
\end{align*}
Therefore, applying the Hajek--Sidak CLT (e.g., Sen and Singer 1994, Theorem 3.3.6, p. 119) and Cramer-Wold device we establish the asymptotic normality of $\pmb{H}^{\top}\pmb{\epsilon}/{\sqrt{n}}$. Next, observe that $\mathbb{V}ar (\pmb{H}^{\top}\pmb{\epsilon}/{\sqrt{n}}) =\sigma^2 (\pmb{H}^{\top} \pmb{H}/n)$ and hence 
\[ 
\sqrt{n}\begin{pmatrix}
\widehat{\bmu}-\bmu\\ \widehat{\pmb\gamma}_s- \pmb{\gamma}_s
\end{pmatrix} \Rightarrow \mathcal{N}(\bzero, \pmb{\Sigma}_{\mu,\gamma_s}),\]
with
\begin{align*}
\pmb{\Sigma}_{\mu,\gamma_s} = \sigma^2 \begin{pmatrix}
\bB^{\top}\pmb{\Xi} \bB & \bB^{\top}\pmb{\Xi} \bC_s \\
\bC_s ^{\top}\pmb{\Xi} \bB & \bC_s ^{\top}\pmb{\Xi} \bC_s 
\end{pmatrix}^{+} \begin{pmatrix}
\bB^{\top}\\ \bC_s ^{\top}
\end{pmatrix}\pmb{\Xi}^3 
(\bB, \bC_s) \begin{pmatrix}
\bB^{\top}\pmb{\Xi} \bB & \bB^{\top}\pmb{\Xi} \bC_s \\
\bC_s ^{\top}\pmb{\Xi} \bB & \bC_s ^{\top}\pmb{\Xi} \bC_s 
\end{pmatrix}^{+}.
\end{align*}
where $\pmb{\Xi}$ is the $|\bnu| \times |\bnu|$ diagonal matrix whose $(i,j)^{th}$ diagonal element is $\theta_{ij}=\lim {n_{ij}/n}$.
 \end{proof}

\subsubsection*{Proof of Theorem \ref{Thm-intermediate-model-gof}:}
\begin{proof}
A bit of algebra shows that  
\begin{align*}
(\overline{\bS}_{\rm ALL}-\widehat{\bnu}_s)=
(\pmb{I}- (\bB,\bC_s)\begin{pmatrix}
\bB^{\top}(\bD/n) \bB & \bB^{\top}(\bD/n) \bC_s \\
\bC_s ^{\top}(\bD/n) \bB & \bC_s ^{\top}(\bD/n) \bC_s 
\end{pmatrix}^{+} \begin{pmatrix}
\bB^{\top}\\ \bC_s ^{\top}
\end{pmatrix}(\bD/n))\overline{\bS}_{\rm ALL}.
\end{align*}
By the CLT we have $\bD^{1/2}(\overline{\bS}_{\rm ALL}- {\bnu}_s) \Rightarrow \mathcal{N} (\bzero, \sigma^2 \pmb{H}_s)$, where $\pmb{H}_s$ is a $|\bnu|\times |\bnu|$ diagonal matrix whose diagonal elements are $1$ if $(i,j)\in\mathcal{E}_n$ and $0$ otherwise. It follows by Slutzky's theorem that $\pmb{U}_{n,\mathcal{S}}=\bD^{1/2}(\overline{\bS}_{\rm ALL}-\widehat{\bnu}_s) \Rightarrow \pmb{U}_{\mathcal{S}}$ where $\pmb{U}_{\mathcal{S}}$ follows a $\mathcal{N}_{|\bnu|}(0,\pmb{\Psi }_{\mathcal{S}})$ and ${\pmb{\Psi }}_{\mathcal{S}}= \sigma^2\, 
\pmb{\Xi}^{1/2} \pmb{M}\,\pmb{\Xi}^{+} \pmb{M}^{\top}\pmb{\Xi}^{1/2}$,
where $$\pmb{M}= \pmb{I} - (\bB,\bC_s)\begin{pmatrix}
\bB^{\top}\pmb{\Xi} \bB & \bB^{\top}\pmb{\Xi} \bC_s \\
\bC_s ^{\top}\pmb{\Xi} \bB & \bC_s ^{\top}\pmb{\Xi} \bC_s 
\end{pmatrix}^{+} \begin{pmatrix}
\bB^{\top}\\ \bC_s ^{\top}
\end{pmatrix}\pmb{\Xi}.$$
Therefore $R_{n,\mathcal{S}} =\pmb{U}_{n,\mathcal{S}}^{\top}\pmb{U}_{n,\mathcal{S}}\Rightarrow \pmb{U}_{\mathcal{S}}^{\top}\pmb{U}_{\mathcal{S}}$. {
Further note that since $\pmb{\Psi}_{\mathcal{S}}$ is symmetric and nonnegative definite we may write its spectral decomposition as $\pmb{\Psi }_{\mathcal{S}}=\textswab{O}^\top\pmb{\Lambda }\textswab{O}$ where $\textswab{O}$ is an orthogonal matrix whose columns are the eigenvectors of $\pmb{\Psi }$ and $\pmb{\Lambda }$ is a diagonal matrix with nonnegative elements which are the eigenvalues of $\pmb{\Psi}_{\mathcal{S}}$. Clearly $\pmb{U}=\pmb{\Psi }_{\mathcal{S}}^{1/2}\bZ$ where $\bZ$ is a $\mathcal{N}_{|\bnu|}(0,\pmb{I})$ RV. It follows that
}
\begin{equation*}
\pmb{U}_{\mathcal{S}}^{\top}\pmb{U}_{\mathcal{S}}= \sum_{i=1}^{{\rm rank}({\pmb{\Psi }}_{\mathcal{S}})}{\lambda}_{i} Z_{i}^{2},
\end{equation*}
where $Z_{1},\ldots ,Z_{t}$ are independent $\mathcal{N}(0,1)$ RVs and ${\lambda} _{1},\ldots ,{\lambda}_{t}$ are the non--zero eigenvalues of the matrix
${\pmb{\Psi }}_{\mathcal{S}}$ which by Remark \ref{rank:psi:lemma} satisfies $t = {\rm rank}({\pmb{\Psi }}_{\mathcal{S}})=\binom{K}{2}-(K+r-1)$.
\end{proof}

{
\begin{remark} \label{rank:psi:lemma}
Algebraically it is difficult to find rank of matrix $\pmb{\Psi}_{\mathcal{S}}$. However, the rank of $\pmb{\Psi}_{\mathcal{S}}$ can be computed using a statistical argument. To do so suppose that the errors $\epsilon_{ijk}$ are IID $\mathcal{N}(0,\sigma^2)$ RVs. It is well known that the LRT for testing $H_{0}:\bnu\in\mathcal{L}\cup \mathcal{C}_{\mathcal{S}}$ versus $H_{1}:\bnu\notin\mathcal{L}\cup \mathcal{C}_{\mathcal{S}}$ follows, under the null a chi--square distribution with $|\mathcal{E}_n|-(K+r-1)$ degrees of freedom. Furthermore, the corresponding Wald test is of the form $\pmb{U}_n^{\top}{\pmb{\Psi }}_n^{+}\pmb{U}_n$ and has the same asymptotic distribution as the LRT and consequently by Brown and Cribari-Neto (1992) the matrix ${\pmb{\Psi }}_n$ must have rank $|\mathcal{E}_n|-(K+r-1)$ for all large $n$ and therefore also in the limit. This established the claim above. 
\end{remark}
}

\bigskip

Next, the non--null distribution of $R_n$ under local alternatives is given.

\begin{proposition} \label{prop:dist:alt:Rn:Wn}
Assume that $\bnu\in \mathcal{S}$ and the condition of Theorem \ref{thm:mu:lambda:estimation:properties} hold. Suppose $\bnu_n=\bnu+n^{-1/2}\pmb{\delta}$ where $\bnu\in\mathcal{S}$ and $\pmb{\delta}\notin \mathcal{S}$ is a fixed vector. If $n\rightarrow \infty$ then
\begin{align*}
R_{n} \Rightarrow \sum_{i=1}^{t}\lambda
_{i}(Z_{i}+\gamma_i)^{2}, 
\end{align*}
where $\lambda_1,\ldots,\lambda_t$ and $Z_1,\ldots,Z_t$ are as in the statement of Theorem \ref{Thm-intermediate-model-gof} and $\gamma_1,\ldots,\gamma_t$ are the elements of the vector $\pmb{\gamma}= \pmb{O} (\pmb{\Psi }_{\mathcal{S}}^{+})^{1/2}\pmb{\Xi}^{1/2}\pmb{\delta}$ which correspond to the nonzero eigenvalues of $\pmb{\Psi}$. Here $\pmb{O}$ is the orthonormal matrix whose columns are the eigenvectors of $\pmb{\Psi}_{\mathcal{S}}$.  
\end{proposition}

\begin{proof}
Under the stated conditions, the $(i,j)^{th}$ entry of $\pmb{U}_{n,\mathcal{S}}$ can be written as
\begin{align*}
U_{n,\mathcal{S},ij}=\sqrt{n_{ij}}(\overline{S}_{ij}-\widehat{\nu }_{\mathcal{S},ij}) 
=\sqrt{n_{ij}}(\overline{S}_{ij}-(\widehat{\nu }_{\mathcal{S},ij}-n^{-1/2}{\delta}_{ij})+ n^{-1/2}{\delta}_{ij})= U_{ij}^* + \sqrt{\frac{n_{ij}}{n}}\,{\delta}_{ij},
\end{align*}
where $ U_{n,ij}^*= \sqrt{n_{ij}}\,(\overline{S}_{ij}-(\widehat{\nu }_{\mathcal{S},ij}-n^{-1/2}{\delta}_{ij}))$. Consequently $\pmb{U}_{n,\mathcal{S}} \Rightarrow \pmb{U}_{\mathcal{S}}$ where $\pmb{U}_{\mathcal{S}}$ is a $\mathcal{N}_{|\bnu|} (\pmb{\Xi}^{1/2}\,\pmb{\delta}, \pmb{\Psi}_{\mathcal{S}})$ RV. It follows that 
\begin{align*}
R_n=&~ \pmb{U}_{n,\mathcal{S}}^{\top} \pmb{U}_{n,\mathcal{S}} \Rightarrow 
\pmb{U}_{\mathcal{S}}^{\top} \pmb{U}_{\mathcal{S}}= 
((\pmb{\Psi }_{\mathcal{S}}^{+})^{1/2}\pmb{U}_{\mathcal{S}}
)^\top\pmb{\Psi}_{\mathcal{S}}((\pmb{\Psi }_{\mathcal{S}}^{+})^{1/2}\pmb{U}_{\mathcal{S}})= ((\pmb{\Psi }_{\mathcal{S}}^{+})^{1/2}\pmb{U}_{\mathcal{S}}
)^\top\textswab{O}^{\top}\pmb{\Lambda}\textswab{O} ((\pmb{\Psi }_{\mathcal{S}}^{+})^{1/2}\pmb{U}_{\mathcal{S}})\\
=&~ \pmb{V}_{\mathcal{S}}^{\top} \pmb{\Lambda} \pmb{V}_{\mathcal{S}},
\end{align*}
where $\pmb{V}_{\mathcal{S}}= \textswab{O} (\pmb{\Psi }_{\mathcal{S}}^{+})^{1/2}\pmb{U}_{\mathcal{S}}$ is a $\mathcal{N}(\pmb{\gamma}, \pmb{H})$ RV, where $\pmb{\gamma}= \textswab{O} (\pmb{\Psi }_{\mathcal{S}}^{+})^{1/2}\pmb{\Xi}^{1/2}\, \pmb{\delta}$ and $\pmb{H}$ is a $|\bnu|\times |\bnu|$ diagonal matrix whose $(i,j)^{th}$ diagonal elements is $1$ if $(i,j)\in\mathcal{E}_n$ and $0$ otherwise. It immediately follows, as in Theorem \ref{Thm-intermediate-model-gof} that 
$$
R_n= \sum_{i=1}^{t} \lambda_i (Z_i+ \gamma_i)^2,
$$
where $\gamma_i$ is the element of $\pmb{\gamma}$ corresponding with the $i^{th}$ nonzero diagonal element of $\pmb{\Lambda}$ denoted by $\lambda_i$.  
\end{proof}

\subsubsection*{Proof of Theorem \ref{thm:select:triad:convergence}:}

We start with an auxiliary Lemma.

\begin{lemma}\label{thm4.4:lemma}
Let $\mathcal{S}= \rm{span}(\bB,\bC_s)$ for some matrix $\bC_s$ satisfying $\rm{span}(\bC_s)\subset \rm{span}(\bC)$. Consider testing $H_0:\bnu \in\mathcal{S}$ versus $H_1:\bnu \notin \mathcal{S}$ using the statistic $R_{n,\mathcal{S}}$ defined in Theorem \ref{Thm-intermediate-model-gof}. Assuming $\bnu \notin \mathcal{S}$, $n_{ij}\rightarrow\infty$ for all $1\leq i<j\leq K$ while $n_{ij}=O(n)$, we then have
\begin{align} \label{eq.power1}
\mathbb{P}(R_{n,\mathcal{S}}< c_{\alpha}) \rightarrow 0  
\end{align}
where $c_{\alpha}=\mathbb{P}_{H_0}(R_{n,\mathcal{S}} \ge c_{\alpha})$ is the $\alpha$ level critical value for $R_{n,\mathcal{S}}$. If, in addition, the errors are subgaussian then for all large $n$ there exists constants $C_1,C_2>0$ such that 
\begin{align} \label{eq.power2}
\mathbb{P}(R_{n,\mathcal{S}}< c_{\alpha}) \leq C_1 \exp(-nC_2).
\end{align}
\end{lemma}

\begin{remark}
Equation \eqref{eq.power1} shows that under the alternative the power of tests for lack of fit approaches unity as the sample size increases to $\infty$. In fact, this assertion has already been established see the text following the statement of Theorem \ref{Thm-intermediate-model-gof}. Therefore we only prove \eqref{eq.power2}.  
\end{remark}

\begin{proof}
Observe that
\begin{equation} \label{eq:R_nS:expanded}
R_{n,\mathcal{S}}=\sum_{1\leq i<j\leq K} {U}_{n,\mathcal{S},ij}^2= \sum_{1\leq i<j\leq K} n_{ij}(\overline{S}_{{\rm ALL},ij}-\widehat{\nu}_{s,ij})^2,
\end{equation}
where $\overline{S}_{{\rm ALL},ij}$ and $\widehat{\nu}_{s,ij}$ are 
$ij^{th}$ elements of $\overline{\bS}_{\rm ALL}$ and $\widehat{\bnu}_s$, respectively. Notice that both $\overline{S}_{{\rm ALL},ij}$ and $\widehat{\nu}_{s,ij}$ are linear functions of the errors. Therefore $(\overline{S}_{{\rm ALL},ij}-\widehat{\nu}_{s,ij})$ is also a subgaussian random variable with mean $\delta_{ij}$ and parameter $\tau_{ij}^2\leq \infty$. Note that under the null, i.e., when $\bnu \in \mathcal{S}$, we have $\delta_{ij}=\mathbb{E}(\overline{S}_{{\rm ALL},ij}-\widehat{\nu}_{s,ij})=0$ for all pairs $(i,j)$. However, under the alternative $\delta_{ij} \neq 0$ for some pairs $(i,j)$.   

Continuing,
\begin{align*}
\mathbb{P}(|\sqrt{n_{ij}}(\overline{S}_{{\rm ALL},ij}-\widehat{\nu}_{s,ij})- \sqrt{n_{ij}}\delta_{ij}|\geq \epsilon)\leq 2 \exp(-\epsilon^2n_{ij}/2\tau_{ij}^2)
\end{align*}
In other words
\begin{align*}
\mathbb{P}(\sqrt{n_{ij}}(\overline{S}_{{\rm ALL},ij}-\widehat{\nu}_{s,ij})\notin(\sqrt{n_{ij}}\delta_{ij}-\epsilon, \sqrt{n_{ij}}\delta_{ij}+\epsilon))\leq 2 \exp(-\epsilon^2n_{ij}/2\tau_{ij}^2),
\end{align*}
so, for all large $n$
\begin{align*}
\mathbb{P}({n_{ij}}(\overline{S}_{{\rm ALL},ij}-\widehat{\nu}_{s,ij})^2\notin({n_{ij}}\delta_{ij}^2-\epsilon_{ij}, {n_{ij}}\delta_{ij}^2+\epsilon_{ij}))\leq 2 \exp(-\epsilon^2n_{ij}/2\tau_{ij}^2)\\    
\end{align*}
for some $\epsilon_{ij}>0$ where $\epsilon_{ij}\rightarrow 0$ when 
$\epsilon\rightarrow 0$. Next observe that 
$$\cup_{1\leq i<j\leq K}\{{U}_{n,\mathcal{S},ij}^2\in (n_{ij}\delta_{ij}^2-\epsilon_{ij}, n_{ij}\delta_{ij}^2+\epsilon_{ij})\}\subset \{R_{n,\mathcal{S}}\in (\sum_{1\leq i<j\leq K}n_{ij}\delta_{ij}^2-\widetilde{\epsilon}, \sum_{1\leq i<j\leq K}n_{ij}\delta_{ij}^2+\widetilde{\epsilon})\}$$
where $\widetilde{\epsilon}= \sum_{1\leq i<j\leq K}\epsilon_{ij}$. 
It follows that for all large $n$
\begin{align*}
&~ \mathbb{P}(R_{n,\mathcal{S}}\notin (\sum_{1\leq i<j\leq K}n_{ij}\delta_{ij}^2-\widetilde{\epsilon}, \sum_{1\leq i<j\leq K}n_{ij}\delta_{ij}^2+\widetilde{\epsilon}))\\
\leq&~ \mathbb{P}(\cup_{1\leq i<j\leq K}\{{U}_{n,\mathcal{S},ij}^2\notin (n_{ij}\delta_{ij}^2-\epsilon_{ij}, n_{ij}\delta_{ij}^2+\epsilon_{ij})\})
\leq \sum_{1\leq i<j\leq K} 2 \exp(-\epsilon^2n_{ij}/2\tau_{ij}^2)\\
\leq&~ C_1 \exp(-n\epsilon^2C_2),
\end{align*}
for $C_1=2\binom{K}{2}$ and $C_2=\min\{n_{ij}/(2n\tau_{ij}^2): 1\leq i<j\leq K\}$. 
Therefore 
\begin{align*}
&~ \mathbb{P}(R_{n,\mathcal{S}} \in (\sum_{1\leq i<j\leq K}n_{ij}\delta_{ij}^2-\widetilde{\epsilon}, \sum_{1\leq i<j\leq K}n_{ij}\delta_{ij}^2+\widetilde{\epsilon}))\geq 1- C_1 \exp(-n\epsilon^2C_2).
\end{align*}
However for any $\epsilon$ and all large $n$, $\sum_{1\leq i<j\leq K}n_{ij}\delta_{ij}^2-\widetilde{\epsilon}>c_{\alpha}$ so for all large $n$
\begin{align*}
\mathbb{P}(R_{n,\mathcal{S}}\geq c_{\alpha}) \geq 1- C_1 \exp(-nC_2).
\end{align*}
Taking complements completes the proof.
\end{proof}

\begin{remark}
Lemma \ref{thm4.4:lemma} shows that under subgaussian errors the probability of not rejecting the null when applying ${\rm FTBS}$ it is false is exponentially small. 
\end{remark}

\noindent We now continue with the proof of Theorem \ref{thm:select:triad:convergence}.

\medskip

\begin{proof}
FTBS is applied by $(i)$ testing $\binom{K}{2}$ hypotheses $H_0:\nu_{{\rm cyclic},ij}=0$, ${1\leq i<j\leq K}$ and generating an estimated tick--table $\bC_{\bnu_{\rm cyclic}^{*}}$; and $(ii)$ then sequentially testing for lack of fit using Theorem \ref{Thm-intermediate-model-gof}. At most $3$ such tests are applied. The set of selected triads obtained after applying $(i)$ and $(ii)$ above is denoted by $\widehat{\mathcal{S}}$. 

By Theorem \ref{thm:ticks:real:model} there exists a minimal model $\mathcal{S}$ in which all triads have at least one tick in $\bC_{\bnu_{\rm cycllic}}$. Further note that $\mathcal{S}\subseteq\mathcal{S}_b$ for some $b\in\{1,\ldots,4\}$. Now, if there is no cyclicality in $\mathcal{S}$ then $\mathcal{S}\subseteq\mathcal{S}_1$; if all triads in $\mathcal{S}$ have three ticks then $\mathcal{S}\subseteq\mathcal{S}_2$; if some triads in $\mathcal{S}$ have two ticks then $\mathcal{S}\subseteq\mathcal{S}_3$; and if some triads in $\mathcal{S}$ have one tick then $\mathcal{S}\subseteq\mathcal{S}_4$. Thus if $\bC_{\bnu_{\rm cycllic}}$ was known then applying Theorem \ref{Thm-intermediate-model-gof} would guarantee that FTBS would select $\widehat{\mathcal{S}}\supset \mathcal{S}$. 

Now, the event $\{\bnu_{\rm cyclic}\notin {\rm span}(\widehat{\mathcal{S}})\}$ will occur if either
$\{\bC_{\bnu_{\rm cycllic}} \not\subset \bC_{\bnu_{\rm cycllic}^*}\}$ or if for some $a\in\{1,\ldots,3\}$ smaller than $b$ the event $\{R_{n,\mathcal{S}_a}<c_{\alpha,a} \}$ occurs. Also note that if $\{\bC_{\bnu_{\rm cycllic}} \subset \bC_{\bnu_{\rm cycllic}^*}\}$ then $\mathcal{S}_a\subset \mathcal{S}_a^*$, moreover $\mathcal{S}_b\subset \mathcal{S}_a^*$ may hold for some $b>a$. It follows that: 
\begin{align} 
\mathbb{P}(\bnu_{\rm cyclic}\notin {\rm span}(\widehat{\mathcal{S}}))=&~ \mathbb{P}(\{{\bC}_{{\bnu}_{\rm cyclic}}\not\subset {\bC}_{{\bnu}_{\rm cyclic}^*}\} \cup_{\{ a:\, \mathcal{S}_b\not\subset \mathcal{S}_a^*\}}(\{R_{n,\mathcal{S}_a}<c_{\alpha,a} \}\cap \{{\bC}_{{\bnu}_{\rm cyclic}}\subset {\bC}_{{\bnu}_{\rm cyclic}^*}\})))\nonumber\\
\leq&~ \mathbb{P}({\bC}_{{\bnu}_{\rm cyclic}}\not\subset {\bC}_{{\bnu}_{\rm cyclic}^*})+ \sum_{\{ a:\, \mathcal{S}_b\not\subset \mathcal{S}_a^*\}}\mathbb{P}(\{R_{n,\mathcal{S}_a^*}<c_{\alpha,a} \}). \label{eq:prob:false:select}
\end{align}
By Lemma \ref{thm4.4:lemma} if $\mathcal{S}_b\not\subset \mathcal{S}_a^*$ then with high probability we will reject the hypothesis that $\mathcal{S}\subset \mathcal{S}_a^*$. Therefore the sum in \eqref{eq:prob:false:select} is over set $\{ a:\mathcal{S}_b\not\subset \mathcal{S}_a^*\}$.

Next we bound the individual probabilities in \eqref{eq:prob:false:select}. First note that if $\nu_{{\rm cyclic},ij} \neq 0$ and the procedure described in Remark \ref{remark:4.1} is followed then $\mathbb{P}(\phi_{ij}(\widehat\bnu_{{\rm cyclic}}=1) \rightarrow 1$, i.e., the null $H_0:\nu_{{\rm cyclic},ij}=0$ is rejected with probability tending to one. Since $K$ is finite all elements of $\bnu_{\rm cyclic}$ which are different than $0$ will be identified with probability tending to one. Furthermore approximately $\alpha(\binom{K}{2}-\sum_{1\leq i<j\leq K}\mathbb{I}({\nu_{{\rm cyclic},ij}}\neq0))$ other elements of $\bnu$ will be falsely flagged as non--zero. We have just proved that
\begin{align} \label{eq:equal:tick:table}
\mathbb{P}({\bC}_{{\bnu}_{\rm cyclic}}\not\subset {\bC}_{{\bnu}_{\rm cyclic}^*})\to 0.
\end{align}
Consequently $\mathcal{S}_a\subset \mathcal{S}_a^*$ for $a\in\{1,\ldots,4 \}$ with probability tending to $1$ as $n\to\infty$. Next using Lemma \ref{thm4.4:lemma} for $\bnu\in\mathcal{S}_b$ and $b>a$, we have
\begin{align} \label{eq:triad:select:fts:prob:limit}
\mathbb{P}(\{R_{n,\mathcal{S}_a^*}<c_{\alpha,a} \})\to 0.
\end{align}
Therefore \eqref{eq:equal:tick:table} and \eqref{eq:triad:select:fts:prob:limit} imply that
\begin{align*} 
\mathbb{P}(\bnu_{\rm cyclic}\notin {\rm span}(\widehat{\mathcal{S}})) \to 0,
\end{align*}
establishing the first stated claim. 

Next, recall that $\widehat{\bnu}_{\rm cyclic}= \bC \widehat{\pmb \gamma}$ where $\widehat{\pmb \gamma}=\pmb{M}(\bnu+\pmb{\epsilon})$ where the matrix $\pmb{M}$ can be deduced from Equation \eqref{nu:thm4.1}. Thus for any pair $(i,j)$ the random variable $\widehat{\nu}_{{\rm cyclic},ij}$ is a linear combination of the errors. Since the errors are IID subgaussian RVs (with parameter $\tau^2\geq \sigma^2$) so is $\widehat{\nu}_{{\rm cyclic},ij}$. In fact by Remark \ref{remark:4.1}, $\sqrt{n}(\widehat{\bnu}_{\rm cyclic}- \bnu_{\rm cyclic}) \Rightarrow \mathcal{N}(\bzero, \sigma^2 \textswab{C}\,\pmb{\Xi}^{+}\textswab{C}^{\top})$, thus for every $n$ the random varaible $\widehat\nu_{{\rm cyclic},ij}- \nu_{{\rm cyclic},ij}$ is a zero mean subgaussian with parameter ${\tau}_{ij}^2/\sqrt{n}$, where ${\tau}_{ij}^2\geq {\sigma}^2_{{\rm cyclic},ij}$ and ${\sigma}^2_{{\rm cyclic},ij}$ is $(i,j)^{th}$ diagonal element of $\sigma^2 \textswab{C}\,\pmb{\Xi}^{+}\textswab{C}^{\top}$. It follows that
\begin{align} \label{prob_bound:nuij_nonzero}
\mathbb{P}(|\widehat\nu_{{\rm cyclic},ij}- \nu_{{\rm cyclic},ij}| \geq \epsilon)\leq 2 e^{-\epsilon^2/({\tau}_{ij}^2/n)}= 2 e^{-\epsilon^2n/{\tau}_{ij}^2}.
\end{align}
Next using union bound and \eqref{prob_bound:nuij_nonzero} we have 
\begin{align*} 
\mathbb{P}(\cup_{1\leq i<j\leq K} \{|\widehat\nu_{{\rm cyclic},ij}- \nu_{{\rm cyclic},ij}| \geq \epsilon \})
&~\leq \sum_{1\leq i<j\leq K} \mathbb{P}( \{|\widehat\nu_{{\rm cyclic},ij}- \nu_{{\rm cyclic},ij}| \geq \epsilon \})\\
&~\leq \sum_{1\leq i<j\leq K} 2 e^{-\epsilon^2n/{\tau}_{ij}^2}.
\end{align*}
Let ${\tau}_{{\rm cyclic}}^2= \min\{{\tau}_{ij}^2: 1\leq i<j\leq K\}$. Then
\begin{align*}
\mathbb{P}(\cup_{1\leq i<j\leq K} \{|\widehat\nu_{{\rm cyclic},ij}- \nu_{{\rm cyclic},ij}| \leq \epsilon \}) 
\geq 1- \sum_{1\leq i<j\leq K} 2 e^{-\epsilon^2n/{\tau}_{ij}^2}
\geq 1- \binom{K}{2} 2 e^{-\epsilon^2n/{\tau}_{{\rm cyclic}}^2}.
\end{align*}
Since $K$ is finite, there exist $\bar{C}_{1},\bar{C}_{2}> 0$ such that 
$$\mathbb{P}(\cup_{1\leq i<j\leq K} \{|\widehat\nu_{{\rm cyclic},ij}- \nu_{{\rm cyclic},ij}| \geq \epsilon \})\leq \bar{C}_{1} e^{(-n\bar{C}_{2})}.$$
Therefore not rejecting $H_0:\nu_{{\rm cyclic},ij}=0$ when $H_0$ is false has probability
\begin{align} \label{eq:equal:tick:table:prob:bound}
\mathbb{P}({\bC}_{{\bnu}_{\rm cyclic}}\not\subset {\bC}_{{\bnu}_{\rm cyclic}^*})\leq \bar{C}_{1} e^{(-n\bar{C}_{2})}.
\end{align}
Next using Lemma \ref{thm4.4:lemma} for $\bnu\in\mathcal{S}_b$ and $b>a$, we have
\begin{align} \label{eq:triad:select:fts:prob}
\mathbb{P}(\{R_{n,\mathcal{S}_a^*}<c_{\alpha,a} \})\leq C_{a1} e^{(-nC_{a2} )}.
\end{align}
Thus \eqref{eq:prob:false:select}, \eqref{eq:equal:tick:table:prob:bound} and \eqref{eq:triad:select:fts:prob} imply that
\begin{align*} 
\mathbb{P}(\bnu_{\rm cyclic}\notin {\rm span}(\widehat{\mathcal{S}})) \leq&~ \mathbb{P}({\bC}_{{\bnu}_{\rm cyclic}}\not\subset {\bC}_{{\bnu}_{\rm cyclic}^*})+ \sum_{1\leq a<b}\mathbb{P}(\{R_{n,\mathcal{S}_a^*}<c_{\alpha,a} \})\\
\leq&~ \bar{C}_{1} e^{(-n\bar{C}_{2})} + \sum_{1\leq a<b} C_{a1} e^{(-nC_{a2} )}\leq C_1 e^{(-nC_2 )}
\end{align*}
for some finite $C_1,C_2>0$. Therefore there exists $C_1,C_2>0$ such that
\begin{align*} 
\mathbb{P}(\bnu_{\rm cyclic}\in {\rm span}(\widehat{\mathcal{S}}))\geq 1- C_1 e^{(-nC_2 )}.
\end{align*}
Thus establishing the second claim. 

By Theorem \ref{thm:ticks:real:model} if triads in the minimal model share at most index then the minimal model is unique, moreover only triads which are part of the minimal model will have three ticks. Condition \eqref{eq.alpha.pi} guarantees that 
$H_0:{\nu}_{{\rm cyclic},ij}=0$ is rejected with probability tending to one if and only if ${\nu}_{{\rm cyclic},ij}\neq 0$. Thus $\mathbb{P}({\bC}_{{\bnu}_{\rm cyclic}}= {\bC}_{{\bnu}_{\rm cyclic}^*})\to 1$. Further note that under the stated condition $\mathcal{S}=\mathcal{S}_2$ and $\mathcal{S}_2=\mathcal{S}_2^*$ with probability tending to $1$. Since the hypotheses in Step 2 of FTBS are tested at level $\alpha_n\rightarrow0$ the algorithm FTBS will terminate at $a=2$ with probability tending to one. In other words $\mathbb{P}(\mathcal{S}=\widehat{\mathcal{S}})\to 1$, concluding the proof. 
\end{proof}

{\section{The relationship between the reduced and full model} \label{appendix:relation:reduced:full}

Section B explore the relationship between the full and reduced model. In particular we investigate what occurs when model \eqref{model.Y_ijk} is fit under the assumption that $\bnu \in \mathcal{L}$, in circumstances where this assumption  may not hold.

\begin{theorem} \label{cyclicality:lemma:nu:mu}
Assume the errors are IID $\mathcal{N}(0,\sigma^2)$ RVs. Let $\mathrm{KL}(\mathcal{N},\mathcal{L})$ denote the Kullback--Leibler divergence between the distribution of $\bY$ indexed by $\bnu\in\mathcal{N}$ and $\bnu\in\mathcal{L}$ respectively. Then: 
\begin{align*}
{\bmu} =\arg\min_{\bmu\in\mathbb{R}^K}\mathrm{KL}(\mathcal{N},\mathcal{L})
= \arg\min_{\bmu\in\mathbb{R}^K}\sum_{1\leq i<j\leq K}n_{ij}(\nu_{ij}-(\mu_{i}-\mu_{j}))^{2}.
\end{align*}
Moreover, if there exists a spanning tree $\mathcal{T}\subset\mathcal{G}$ such that $\min \{n_{ij}:\left( i,j\right) \in \mathcal{T} \}\rightarrow \infty $ as $n\rightarrow \infty$, then the LSE of the reduced model, given by $\widehat{\bmu}=\bN^+ \bS$, will converge to
\begin{align}\label{mu:nu:kldiv}
\arg\min_{\bmu\in\mathbb{R}^K}\{\sum_{1\leq i<j\leq K}\theta_{ij}(\nu_{ij}-( \mu _{i}-\mu _{j}) )^{2}: \pmb{v}^{\top}\bmu=0\}. 
\end{align}
The minimiser of \eqref{mu:nu:kldiv} is unique if there exists a spanning tree $\mathcal{T}\subset \mathcal{G}$ such that
\begin{equation*}
\min \{n_{ij}:(i,j)\in \mathcal{T}\}/n\to c\in(0,\infty)\text{ as }n\to\infty.
\end{equation*}
Furthermore, the solution of \eqref{mu:nu:kldiv} is independent of $\pmb{\Theta}$ if and only if $\bnu \in \mathcal{L}$. Finally if the constraint $\pmb{v}=\b1$ is imposed then the linear relation 
\begin{align} \label{nu:mu:relation}
\bmu = (\bB^{\top} \pmb{\Xi} \bB)^{+} \bB^{\top} \pmb{\Xi} \bnu.
\end{align}
holds.  
\end{theorem}

\begin{proof}
By assumption $\bY\sim {\mathcal{N}}_n( \bL\bnu,\sigma^2\pmb{I})$, where matrix $\bL$ is defined in Section \ref{section:complete:PCGs}. Moreover if $\bnu\in\mathcal{L}$ then $\nu_{ij}=\mu_i-\mu_j$ and for brevity will be denoted it by $\bnu_{\mu}$. Let $\pmb{f}_1$ and $\pmb{f}_2$ be the densities corresponding to $\bnu\in\mathcal{N}$ and $\bnu\in\mathcal{L}$, respectively. The Kullback--Leibler divergence between models indexed by $\bnu\in\mathcal{N}$ and $\bnu\in\mathcal{L}$ is
\begin{align*}
\mathrm{KL}(\mathcal{N},\mathcal{L}) = &~
\mathbb{E}_{\pmb{f}_1}(\log{\pmb{f}_1}-\log{\pmb{f}_2})
= \frac{1}{2\sigma^2}\mathbb{E}_{\pmb{f}_1}(-(\bY-\bL\bnu)^{\top}(\bY-\bL\bnu) + (\bY-\bL\bnu_{\mu})^{\top}(\bY-\bL\bnu_{\mu}))\\
= &~  \frac{1}{2\sigma^2}\mathbb{E}_{\pmb{f}_1}( 2(\bY-\bL\bnu)^{\top} (\bL\bnu-\bL\bnu_{\mu}) + (\bL\bnu-\bL\bnu_{\mu})^{\top}(\bL\bnu-\bL\bnu_{\mu}))\\
= &~ \frac{1}{2\sigma^2} (\bnu-\bnu_{\mu})^{\top}\bL^{\top}\bL (\bnu-\bnu_{\mu}) =  \frac{1}{2\sigma^2} \sum_{1\leq i<j\leq K}n_{ij}(\nu_{ij}-( \mu _{i}-\mu _{j}) )^{2}.
\end{align*}
It follows that $\mathrm{KL}(\mathcal{N},\mathcal{L})$ is minimized by
\begin{align*}
{\bmu} = \arg\min_{\bmu\in\mathbb{R}^K}\mathrm{KL}(\mathcal{N},\mathcal{L}) = \arg\min_{\bmu\in\mathbb{R}^K}\sum_{1\leq i<j\leq K}n_{ij}(\nu_{ij}-(\mu_{i}-\mu_{j}))^{2}.
\end{align*}

Next observe that 
\begin{align*}
\widehat{\bmu} =&~\arg\min_{\bmu\in\mathbb{R}^K}\frac{1}{n} \sum_{1\leq i<j\leq K}\sum_{t=1}^{n_{ij}}(Y_{ijt}-\left( \mu _{i}-\mu _{j}\right) )^{2}
= \arg\min_{\bmu\in\mathbb{R}^K}\frac{1}{n}\sum_{1\leq i<j\leq K}\sum_{t=1}^{n_{ij}} (\nu_{ij}-(\mu _{i}-\mu _{j}) +\epsilon_{ijt})^{2}\\
=&~\arg\min_{\bmu\in\mathbb{R}^K} \sum_{1\leq i<j\leq K} \frac{n_{ij}}{n} (\nu_{ij}-(\mu_{i}-\mu_{j}))^{2} - 2 \sum_{1\leq i<j\leq K}\frac{n_{ij}}{n}(\nu_{ij}-(\mu _{i}-\mu _{j})) \frac{1}{n_{ij}}\sum_{t=1}^{n_{ij}} \epsilon_{ijt} \\
&~+\frac{1}{n} \sum_{1\leq i<j\leq K}\sum_{t=1}^{n_{ij}} \epsilon_{ijt}^2. 
\end{align*}
By the law of large numbers the third term above converges to $\sigma^2$ which is independent of $\bmu$. By assumption  $n_{ij}/n=\theta_{ij} + o(1)$ and consequently we can replace $n_{ij}/n$ appearing in the first term above with $\theta_{ij} + o(1)$. If $\theta_{ij}=0$ then the second term is $o_p(1)$ whereas if $\theta_{ij}>0$ then ${1}/{n_{ij}}\sum_{t=1}^{n_{ij}} \epsilon_{ijt}\to 0$; hence in both cases the second term is $o_p(1)$. Putting it all together we find that  
\begin{align*}
\widehat\bmu= \arg\min_{\bmu\in\mathbb{R}^K} \sum_{1\leq i<j\leq K}\theta_{ij}(\nu_{ij}-(\mu_{i}-\mu_{j}))^{2} + o_p(1).
\end{align*}
establishing \eqref{mu:nu:kldiv}. Further note that
\begin{align} \label{eq.QF}
\sum_{1\leq i<j\leq K}\theta_{ij}(\nu_{ij}-(\mu_{i}-\mu_{j}))^{2} = (\bnu - \bB\bmu)^{\top} \pmb{\Xi} (\bnu - \bB\bmu).
\end{align}
By assumption there exist a tree $\mathcal{T}$ such that $\theta_{ij}>0$ for all $(i,j)\in\mathcal{T}$; clearly a unique minimizer exist on $\mathcal{T}$ and consequently on every $\mathcal{G} \supset \mathcal{T}$. For a finite sample analogue see Theorem 2.1 in Singh et al. (2025). Further note that if  $\bnu \in \mathcal{L}$ then \eqref{mu:nu:kldiv} is minimized by choosing $\bmu$ such that $\nu_{ij}=\mu_i-\mu_j$ in which case \eqref{eq.QF} is identically zero for all possible choices of $\pmb{\Theta}$. If however $\bnu \notin \mathcal{L}$ then \eqref{eq.QF} is always a function of $\pmb{\Theta}$. The Larnagian associated with \eqref{mu:nu:kldiv}
is 
\begin{align*}
L = (\bnu - \bB\bmu)^{\top} \pmb{\Xi} (\bnu - \bB\bmu) + \lambda \pmb{v}^{\top}\bmu,
\end{align*}
where $\b1^{\top}\pmb{v}\neq0$. Note that gradient of $L$ is $\nabla L =(\nabla_{\bmu}L, \nabla_{\lambda}L)$ where 
\newline $\nabla_{\bmu}L = - \bB^{\top} \pmb{\Xi}\bnu + \bB^{\top} \pmb{\Xi} \bB \bmu+ \lambda\pmb{v}/2$ and $\nabla_{\lambda}L = \pmb{v}^{\top}\bmu$. Following the proof of Theorem 2.1 in Singh et al. (2025) it can be shown that 
\begin{align*}
\widehat\bmu= (\bB^{\top} \pmb{\Xi} \bB)^{+}\bB^{\top} \pmb{\Xi}\bnu + [\pmb{I}- (\bB^{\top} \pmb{\Xi} \bB)^{+}\bB^{\top} \pmb{\Xi} \bB ]\pmb{\alpha},
\end{align*}
where $\pmb \alpha\in\mathbb{R}^K$ is arbitrary which reduces to
\begin{align} \label{hat:mu}
\widehat{\bmu}= (\bB^{\top} \pmb{\Xi} \bB)^{+} \bB^{\top} \pmb{\Xi} \bnu
\end{align}
when $\pmb{v}=\b1$.
\end{proof}

\bigskip

Theorem \ref{cyclicality:lemma:nu:mu} explores the relationship between $\bmu\in \mathbb{R}^K$ and $\bnu\in\mathbb{R}^{K(K-1)/2}$. One of its consequences is that the LSE, assuming the reduce model, given by $\widehat{\bmu}=\bN^+ \bS$, will converge to the LHS of \eqref{mu:nu:kldiv}. Obviously this limit depends very strongly on the limiting value of the Laplacian, as illustrated by the following example. 

\begin{example} \label{different:graph:nu:mu}
Let $K=4$ and suppose that $\bnu = (-2,0,2,0,2,0)$. It is easy to check that $\bnu \notin \mathcal{L}$. By Theorem \ref{cyclicality:lemma:nu:mu} the LSE of $\bmu$, will converge to the RHS of \eqref{nu:mu:relation}. If the comparison graph is complete and balanced, i.e., $n_{ij}=m$ for $1\leq i<j\leq K$, then $\pmb{\Xi}=diag(1/6,1/6, 1/6, 1/6, 1/6,1/6)$; whereas if it is a balanced path graph, i.e., $n_{12}=n_{23}=n_{34}=m$, then $\pmb{\Xi}=diag(1/3,0,0,1/3,0,1/3)$. In both cases $\bB$ is given by 
\begin{align*} 
\pmb{B}^{\top} = 
\begin{pmatrix}
%12 & 13 & 14 & 23 & 24 & 34\\
1 & 1 & 1 & 0 & 0 & 0\\
-1 & 0 & 0 & 1 & 1 & 0\\
0 & -1 & 0 & -1 & 0 & 1\\
0 & 0 & -1 & 0 & -1 & -1
\end{pmatrix}.
\end{align*}
Using \eqref{nu:mu:relation} we find that the the LSE will converge to $(0,1,0,-1)$ for the complete graph and to $(-1.5, 0.5,0.5,0.5)$ for the path graph. The resulting merits are obviously very different and topology dependent.  
\end{example}

It is worth noting that if we substitute the sample versions of $\pmb{\Xi}$ and $\bnu$ on the RHS of  \eqref{nu:mu:relation} then its LHS, i.e., $\bmu$, coincides with the LSE $\widehat{\bmu}=\bN^+ \bS$. Therefore  \eqref{nu:mu:relation} provides an alternative way of computing the LSE. Further note that using \eqref{nu:mu:relation} one can find the value of $\bnu \in \mathcal{L}$, denoted by $\bnu_{\bmu}$, providing the best fit for the model when $\bnu \in \mathcal{N}$. In fact  
\begin{align*}
\bnu_{\bmu} = \bB \bmu = \bB (\bB^{\top} \pmb{\Xi} \bB)^{+} \bB^{\top} \pmb{\Xi} \bnu
\end{align*}
so $\|\bnu -\bnu_{\bmu}\| \leq \|\pmb{I}- \bB (\bB^{\top} \pmb{\Xi} \bB)^{+} \bB^{\top} \pmb{\Xi}\|_F \|\bnu\|$, where $\|\cdot\|_F$ denotes the Frobenius norm. It should also be emphasized that $\bnu_{\bmu}$ is a projection onto $\mathcal{L}$ if and only if $\pmb{\Xi}= a \pmb{I}$ for some $a\in\mathbb{R}$, i.e., if and only if the comparison graph is balanced in which case $n_{ij}=m$ for all pairs $(i,j)$. 

}

\section{Ranking under Intransitive Models}
\label{section:ranking:intransitivity}

As noted in Section \ref{section:introduction} in the main text, if there are no cyclicalities, i.e., $\bnu\in\mathcal{L}$ then $i \succeq j$ if and only if $\mu_i \ge \mu_j$. As expected, ranking is more complicated when $\bnu\notin\mathcal{L}$. 

\begin{theorem}\label{thm:rank:strong:transitive}
Suppose \eqref{bnu:decomposed:linear:cyclic} holds and let $\bnu$ be transitive. If so $i \succeq j$ implies that $\mu_i \ge \mu_j$ if $\bnu$ is strongly transitive but not necessarily if $\bnu$ is weakly transitive.  
\end{theorem}    

%\subsubsection*{Proof of Theorem \ref{thm:rank:strong:transitive}:}
\begin{proof}
Pick a triplet $(1,2,3)$, say. Recall, see Example \ref{nu:linear:cyclic:example}, that when $K=3$ the relevant means satisfy 
\begin{align*}
\begin{pmatrix}
\nu_{12}\\ \nu_{13} \\ \nu_{23}
\end{pmatrix} =
\begin{pmatrix}
\mu_1-\mu_2+\gamma\\
\mu_1-\mu_3-\gamma\\
\mu_2-\mu_3+\gamma
\end{pmatrix},
\end{align*}
for some $\gamma$. Suppose that $\bnu$ is strongly transitive and without loss of generality assume that $1\succeq 2 \succeq 3$. If so 
\begin{align*} 
\nu_{12},\nu_{23}\geq0 \text{ and }\nu_{13}\geq \max \{\nu_{12},\nu_{23}\}.
\end{align*}
If $\gamma<0$, then $\nu_{12}\geq 0$ implies that $\mu_1 - \mu_2 >- \gamma>0$ so $\mu_1>\mu_2$. Similarly $\nu_{23}\geq 0$ implies $\mu_2>\mu_3$. Together these imply that $\mu_1\geq\mu_2\geq \mu_3$. Next if $\gamma>0$ then strong transitivity implies that $\mu_1 - \mu_3 - \gamma\geq \mu_1 - \mu_2 + \gamma$, so $\mu_2-\mu_3>2\gamma>0$. Similarly $\mu_1 - \mu_3 - \gamma\geq \mu_2 - \mu_3 + \gamma$ implies $\mu_1-\mu_2>2\gamma>0$. Combining these two relations we find that $\mu_1\geq\mu_2\geq \mu_3$ as required. Finally if $\gamma =0$ then we have $\mu_1-\mu_2\geq0$ and $\mu_2-\mu_3\geq0$ which immediately imply that $\mu_1\geq\mu_2\geq \mu_3$. Since this argument holds for any triplet we conclude that strong stochastic transitivity is equivalent to the ordering of the merits.

Next it will be shown that if $\bnu$ is weakly transitive the merits may not be ordered. Under weak transitivity  
$$\nu_{12},\nu_{23}\geq0 \text{ implies that }\nu_{13}\geq 0.$$
Set $\gamma>0$ and $\mu_2 = \mu_1 + \gamma/2$. If so $\nu_{12},\nu_{23},\nu_{13}\geq 0$ holds but $\mu_1<\mu_2$. We conclude that under weak transitivity the relations $1\succeq 2 \succeq 3$ does not imply that $\mu_1\geq \mu_2\geq\mu_3$. This concludes the proof.  
\end{proof}

We conclude that if $\bnu$ is strongly transitive, then the merits can be used for ranking; but not so under weak stochastic transitivity. Nevertheless, items can always be ranked even under weak transitivity. 

\medskip

In practice, the fitted model often produces complex, cyclical preference orderings (as illustrated in Figure \ref{example:preference:graph}). To address situations where a definitive ranking is nonetheless required, some methods are known in literature. Saari (2014) and Saari (2021) addressed this issue by treating cyclicalities as noise. For $i=1,\ldots,K$ he proposed to compute the pseudo merit of the $i^{th}$ item by
\begin{equation} \label{eq.saari.score}
\mu_i^{*} = \sum_{j=1}^{K}  \nu_{ij}
\end{equation}
and rank $i$ above $j$ if the pseudo--merits satisfy $\mu_i^{*} \ge \mu_j^{*}$. This approach is a relative of the well--known row--sum method (Huber, 1963). Further note that for complete and balanced comparison graphs  ranking based on $\mu_i$ and $\mu_i^{*}$ are identical; see the proof of Theorem 5.1 in Singh et al. (2025). In general, the pseudo--merits can be defined by $\sum_{j=1}^{K} \theta_{ij}\nu_{ij}$, where $\theta_{ij}$ is weight given to pair $(i,j)$. In the same spirit one can compute
\begin{equation} \label{eq.dom.score}
\mu_i^{**} = \sum_{j=1}^{K}  \mathbb{I}(\nu_{ij}>0)
\end{equation}
which we refer to as the dominance score. Notice that the dominance score is reminiscent of the well known Borda--Count advocated among some voting theorists (e.g., Saari 2023). Using the dominance score rank $i$ above $j$ if $\mu_i^{**} > \mu_j^{**}$. Ties among the dominance score can be further broken by repeatedly computing \eqref{eq.dom.score} on sets of the form $\{j: \mu_i^{**}=\mu_j^{**}\}$. The following proposition suggests that dominance scores can be used for ranking. 

\begin{proposition}\label{prop:dominance:score:transitive}
If $\mu_i^{**} \neq \mu_j^{**}$ for all $i \neq j$ then $\bnu$ is weakly transitive.
\end{proposition}

%\subsubsection*{Proof of Proposition \ref{prop:dominance:score:transitive}:}
\begin{proof}
Without any loss of generality assume $\mu_1^{**}>\ldots> \mu_K^{**}$. Consequently: (i) $\nu_{1j}>0$ for $2\leq j\leq K$, so $1\succ j$ where $2\leq j\leq K$; (ii) for any $i>1$, $\nu_{ij}>0$ for $j>i$ and $\nu_{ij}<0$ for $j<i$. Therefore for any $i$ items $\{1,\ldots,i-1\}$ are preferred over $i$ and $i$ is preferred over items $\{i+1,\ldots,K\}$. Implying that the preference profile is weakly transitive with $1\succ \ldots \succ K$.
\end{proof}

Equations \eqref{eq.saari.score} and \eqref{eq.dom.score} map $\mathcal{N} \mapsto P_{n}$, where $P_n$ is the space of all permutations of $\{1,\ldots,n\}$. Such mappings are of course not completely satisfactory. In fact, as noted in Section \ref{section:introduction}, cyclicalities are often an important aspect of the data and should not be averaged out and discarded. 

Finally we note that global non--transitivity may allow in some settings for partially transitivity, i.e., the set $\{1, \ldots, K\}$ may be partitioned into $g$ groups satisfying
\begin{align} \label{partition:items}
\langle i_{1,1},\ldots,i_{1,K_1}\rangle \succeq  \langle i_{2,1},\ldots,i_{2,K_2}\rangle \succeq  \ldots \succeq  \langle i_{g,1},\ldots,i_{g,K_g}\rangle,
\end{align}
where $\langle i_{1,1},\ldots,i_{1,K_1}\rangle \succeq  \langle i_{2,1},\ldots,i_{2,K_2}\rangle $ means $i_{1,l_1}\succeq i_{2,l_2}$ for any $1\leq l_1\leq K_1$ and $1\leq l_2\leq K_2$. If strong stochastic transitivity holds among any three items inhabiting three different groups, then by Proposition \ref{thm:rank:strong:transitive} the merits of these items can be used to rank them. If \eqref{partition:items} holds with $g=1$ then $\bnu$ is strongly intransitive although subsets of transitive items exist. A complete ranking in the presence of true cyclicalities is challenging.

\section{Additional Results}
Significant cyclical triads may be identified by testing
\begin{align}\label{H0:cyclic:triad1}
H_{0}:\nu _{ij}+\nu _{jk} +\nu _{ki}= 0
\quad \text{against} \quad
H_{1}:\nu _{ij}+\nu _{jk} +\nu _{ki}\neq 0.
\end{align}
A natural test statistic for \eqref{H0:cyclic:triad1} is
\begin{align} \label{test:cyclicality:3tuples}
T_n = T_n(i,j,k)= \frac{1}{\sqrt{1/n_{ij}+1/n_{jk}+1/n_{ki}}} (\widehat{\nu}_{ij} +\widehat{\nu}_{jk} + \widehat{\nu}_{ki}),
\end{align}
for which:
\begin{proposition} \label{prop:H0:cyclic:triad}
Let $n_{ij},n_{jk}, n_{ik}\to\infty$ at the same rate. Then, under $H_0$, $T_n\Rightarrow \mathcal{N}(0, \sigma^2)$ where $\sigma^2$ can be consistently estimated by $\widehat{\sigma}^2=\sum_{1\leq i<j\leq 
 K}\sum_{k=1}^{n_{ij}}(Y_{ijk}-\widehat{\nu}_{ij})^2/n$. Under local alternatives of the form $\nu _{ij}+\nu _{jk} +\nu _{ki}= \delta\sqrt{1/n_{ij}+1/n_{jk}+1/n_{ki}}$ we have $T_n\Rightarrow \mathcal{N}(\delta, \sigma^2)$. 
\end{proposition} 

%\subsubsection*{Proof of Proposition \ref{prop:H0:cyclic:triad}:}
\begin{proof}
{If $n_{ij}\to\infty$ and errors are IID with zero mean and finite variance $\sigma^2$ then
\begin{align*}
\sqrt{n_{ij}}\,(\widehat{\nu}_{ij}- {\nu}_{ij}) \Rightarrow \mathcal{N}(0, \sigma^2 ).
\end{align*} }
Therefore, under $H_0$, asymptotic normality of $T_n$ is immediate. Next note that $\nu _{ij}+\nu _{jk} +\nu _{ki}= (\bC_{ij}+\bC_{jk}-\bC_{ik}) \pmb{\gamma}$, so under $H_1$, $T_n$ converges in probability to $\delta$ as $n_{ij},n_{jk}, n_{ik}\to\infty$. Consequently using similar algebra again under $H_1$ we get $T_n\Rightarrow \mathcal{N}(\delta, \sigma^2)$ as $n_{ij},n_{jk}, n_{ik}\to\infty$.
\end{proof}

%\newpage
\section{EPL rank sets}
The following table provides the rank sets for the teams participated in EPL season 2022-23.

\begin{table}[!htbp]
\caption{Rank sets} \label{table:rank:sets}
\resizebox{\textwidth}{!}{\begin{tabular}{clll} \hline
{Team} & $S_i$ & $I_i$ & $E_i$ \\ \hline 
1 & 12,13 & 2,3,4,5,6,7,8,9,10,11,14,15,16,17,18,19,20 &  \\
2 & 1,6,7,11,12,13,14,15,19 & 3,4,5,8,10,16,17,18,20 &  9 \\
3 & 1,2,4,5,6,7,10,12,13,14,15,16,18,19,20 & 8,9,11,17 &  \\
4 & 1,2,5,11,12,13 & 3,6,7,8,9,10,14,15,16,17,18,19,20 &  \\
5 & 1,2,13,15 & 3,4,6,7,8,9,10,11,12,14,16,17,18,19,20 &  \\
6 & 1,4,5,10,11,13,14,15,16,19 & 2,3,7,8,9,12,17,18,20 &  \\
7 & 1,4,5,6,8,9,12,13,14,15,17,18,20 & 2,3,10,11,16,19 &  \\
8 & 1,2,3,4,5,6,9,11,12,13,14,15,18,19 & 7,10,16,17,20 &  \\
9 & 1,3,4,5,6,11,12,13,14,15,18,19 & 7,8,10,16,17,20 &  {2} \\
10 & 1,2,4,5,7,8,9,11,12,13,14,15,18,19,20 & 3,6,16,17 &  \\
11 & 1,3,5,7,12,14,15,16,17 & 2,4,6,8,9,10,13,19,20 &  {18} \\
12 & 5,6,13 & 1,2,3,4,7,8,9,10,11,14,15,16,17,18,19,20 &  \\
13 & {11} & 1,2,3,4,5,6,7,8,9,10,12,14,15,16,17,18,19,20 &  \\
14 & 1,4,5,12,13,15,17 & 2,3,6,7,8,9,10,11,16,18,20 &  {19} \\
15 & 1,4,12,13 & 2,3,5,6,7,8,9,10,11,14,16,17,18,19,20 &  \\
16 & 1,2,4,5,7,8,9,10,12,13,14,15,17,18,19 & 3,6,11,20 &  \\
17 & 1,2,3,4,5,6,8,9,10,12,13,15,18,19 & 7,11,14,16,20 &  \\
18 & 1,2,4,5,6,12,13,14,15 & 1,2,4,5,6,12,13,14,15 &  {11} \\
19 & 1,4,5,7,11,12,13,15,18 & 2,3,6,8,9,10,16,17,20 &  {14} \\
20 & 1,2,4,5,6,8,9,11,12,13,14,15,16,17,18,19 & 3,7,10 & \\ \hline
\end{tabular}}
\end{table}

{\setstretch{1}

\end{document}